\documentstyle[epsfig,times]{elsart}
\begin{document}

\newcommand{\re}{\mathop{\mathrm{Re}}}
\newcommand{\im}{\mathop{\mathrm{Im}}}
\newcommand{\D}{\mathop{\mathrm{d}}}
\newcommand{\I}{\mathop{\mathrm{i}}}
\newcommand{\E}{\mathop{\mathrm{e}}}

\noindent {\Large DESY 04-126}

\noindent {\Large July 2004}

\bigskip

\begin{frontmatter}

\journal{Nuclear Instruments and Methods A}
\date{July 13, 2004 \hspace*{3mm} $\mbox{}$}

\title{
A simple method for the determination of the structure of ultrashort
relativistic electron bunches}

\author{E.L.~Saldin},
\author{E.A.~Schneidmiller},
and
\author{M.V.~Yurkov}

\address{Deutsches Elektronen-Synchrotron (DESY),
Hamburg, Germany}

\begin{abstract}

In this paper we propose a new method for measurements of the
longitudinal profile of 100 femtosecond electron bunches for X-ray Free
Electron Lasers (XFELs). The method is simply the combination of two
well-known techniques, which where not previously combined to our
knowledge. We use seed 10-ps  1047 nm quantum laser to produce exact
optical replica of ultrafast electron bunches. The replica is generated
in apparatus which consists of an input undulator (energy modulator),
and the short output undulator (radiator) separated by a dispersion
section. The radiation in the output undulator is excited by the
electron bunch modulated at the optical wavelength and rapidly reaches
100 MW-level peak power. We then use the now-standard method of
ultrashort laser pulse-shape measurement, a tandem combination of
autocorrelator and spectrum (FROG -- frequency resolved optical
gating). The FROG trace of the optical replica of electron bunch
gives accurate and rapid electron bunch shape measurements in a way
similar to a femtosecond oscilloscope. Real-time  single-shot
measurements of the electron bunch structure could provide significant
information about physical mechanisms responsible for generation
ultrashort electron bunches in bunch compressors. The big advantage of
proposed technique is that it can be used to determine the slice energy
spread and emittance in multishot measurements.  It is possible to
measure bunch structure completely, that is to measure peak current,
energy spread and transverse emittance as a function of time. We
illustrate with numerical examples the potential of the proposed method
for electron beam diagnostics at the European X-ray FEL.

\end{abstract}

\end{frontmatter}

\baselineskip 18pt

%
%

\clearpage

\setcounter{page}{1}

\section{Introduction}

The past decade has been tremendous progress in the development of
electron accelerators that produce ultrashort bunches approaching
sub-100 femtosecond durations \cite{ay1,ay2}. The use of ultrashort
electron bunches for both fundamental studies and applications is
increasing rapidly, too \cite{tdr1,tdr-xfel,tdr2}. As electron bunches
shrink in length and grow in utility, the ability to measure them
becomes increasingly important. There are several reasons for this.
First, precise knowledge of the bunch properties is necessary for
verifying theoretical models of bunch creation \cite{fp1,fp2}. Second,
in order to make even shorter bunches, it is necessary to understand
the distortions that limit the length of currently available pulses.
Third, in experiments using these bunches, it is always important to
know at least the pulse length in order to determine the temporal
resolution of a given experiment. Moreover, in many experiments --
studies of X-ray SASE FELs, for example - additional details  of the
bunch structure play an important role in determination of the outcome
of the experiment. Of particular importance is the variation of peak
current, emittance and energy spread along the bunch.  Finally,
numerous applications have emerged for emittance-shaped ultrashort
electron bunches and, of course, it is necessary to be able to measure
the emittance, or energy spread shape of the electron bunch used in
these experiments \cite{emma}.

Measuring ultrashort electron bunches has always been a challenge. For
five years, it was possible to create 100-fs electron bunches, but not
to measure them \cite{ay1,ay2}. Standard electron beam diagnostic tools
are capable to measure bunch charge, projected emittance, and energy
spread of the full electron bunch only. Unfortunately, they fail to
measure the temporal dependence of the charge distribution within the
bunch. It is not possible to measure slice emittance because electron
bunches are so much shorter than the temporal resolution of measurement
devices.  Also, even when projected energy spread of the full electron
bunch  is measured, there is no sufficient information to determine
slice energy spread and energy chirp separately. On the other hand, it
is primarily the slice emittance and slice energy spread of electrons
in axial slices (that are only a small fraction of the full bunch
length) that determine the performance of a X-ray FEL. Thus, there is
an urgent need for development of electron beam instrumentation
allowing to measure bunch structure completely,  that is, to measure
the temporal dependence of the charge, emittance and energy spread
distributions within the bunch.

The new principle of diagnostic techniques described bellow offers a
way for full characterization of ultrashort electron bunches. It is
based on a construction of an exact optical replica of an electron
bunch. The replica synthesizer consists of four elements: the seed
quantum laser, the modulator undulator, dispersion section, and
radiator undulator. The seed laser pulse interacts with electron beam
in the modulator undulator and produces the energy modulation in the
electron bunch. The electron beam then passes through the dispersion
section where the energy modulation is converted to a density
modulation  at the optical wavelength. Particles in a modulated bunch
following a second undulator radiate coherently at a wavelength of the
beam density modulation.  The bandwidth-limited radiation pulse has 10
$\mu$J-level pulse energy. Longitudinal dynamics, in the undulators and
dispersion section is governed by purely single-particle effects where
the results do not depend on the presence of other particles. In
general the radiation field depends on the peak current, local energy
spread and emittance. All steps of the replica synthesis are controlled
by means of the choice of the undulator parameters, dispersion section
strength and value of beta function. The electric field of the wave
radiated in the replica synthesizer with optimized undulator length,
strength of dispersion section and focusing beta function is directly
proportional to the peak current of the electron beam, $E(t) \simeq
{\mathrm{const.}}\times I(t)$, and does not depend on the local energy
spread and emittance. So, measuring electron current profile, $I(t)$,
for a single ultrashort electron pulse is reduced to the problem of a
single-shot, ultrafast laser pulse-shape measurement.

To characterize such short optical pulses, conventional photodetectors
and streak camera detectors do not have fast enough response times.
Special measurement techniques are needed. Early on, it was realized
that the only event fast enough to measure an ultrashort pulse is the
optical pulse itself. A large number of clever schemes have been
developed over the past twenty years to better measure ultrashort laser
pulses. Most of them have been novel experimental implementations and
variations of autocorrelators, but many have also offered additional
information about the pulse, although never full characterization.
Recently there has been a renaissance in this field and several new
techniques have emerged that do achieve full characterization. They
operate, not in the time or frequency domains, but in the
"time-frequency domain." With the most commonly used new
pulse-measurement method, frequency resolved optical gating (FROG), it
is now possible to measure pulses in the visible or IR wavelength
range, pulse lengths and complexities and to do so in manner that is
general, robust, accurate and rigorous \cite{treb}. FROG simply
involves spectrally resolving the signal beam of an intensity
autocorrelator measurement. FROG is a technique to measure ultrashort
laser pulses that optically constructs a spectrogram of a laser pulse.
A two-dimensional (2-D) phase retrieval algorithm is used to extract
the intensity and phase of a pulse from its spectrogram. The algorithm
is fast enough to allow real-time inversion of the FROG spectrograms.
It is also possible to measure the intensity distribution of a single
ultrashort laser pulse. The entire trace can then be obtained on a
single CCD camera image. Recent improvements of the FROG technique have
lead to very sophisticated retrieval procedures, which can rapidly
retrieve the pulse from the FROG trace. Acquisition and reconstruction
rates of up to 10 Hz have been demonstrated, which makes FROG to be an
ideal online tool for aligning complex femtosecond laser systems.

It is clear that the revolution that has taken place only recently in
ultrashort optical pulse measurement has not only yielded powerful new
laser diagnostics, but also has opened up tremendous new possibilities
for ultrafast accelerator technology. With this new-found capability, a
number of otherwise impossible experiments are now possible. We
illustrate with numerical examples the potential of the proposed method
for electron beam diagnostics at the European X-ray FEL. We demonstrate
that the tandem combination of replica synthesizer and new FROG based
femtosecond oscilloscope  can be used to determine the temporal
dependence of the charge distribution within the bunch for single
ultrashort electron bunch. Proposed techniques have emerged that do
achieve full characterization of the ultrashort electron bunches.  The
big advantage of the proposed diagnostic technique is that it can be
used to determine the slice emittance and energy spread for a multishot
measurements. We show that proposed technique can directly obtain the
electron bunch slice energy spread and emittance from data sets of beta
function and dispersion section strength scans.

\section{Full characterization of femtosecond electron bunches by
optical replica measurements}

\subsection{Optical replica synthesis}

\begin{figure}[tb]
\begin{center}
\epsfig{file=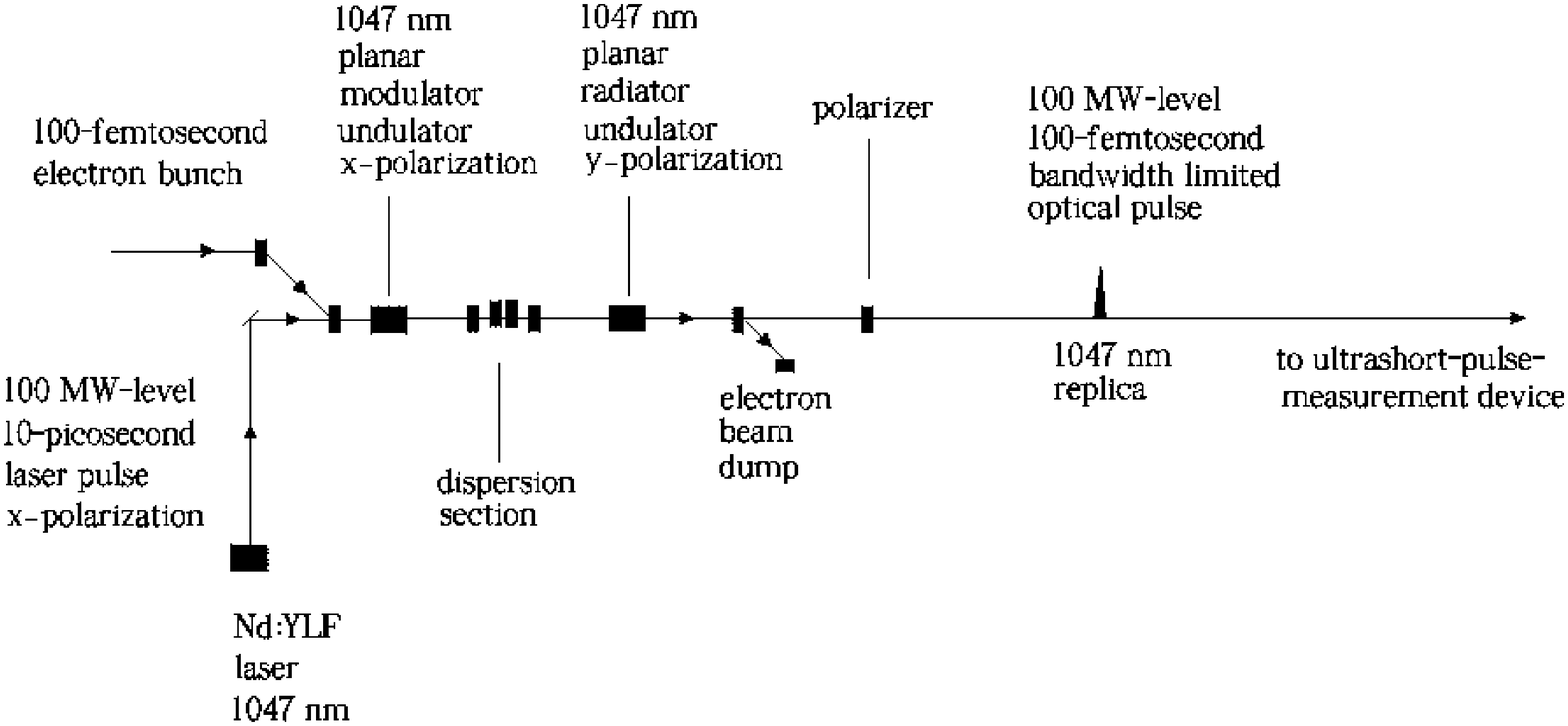,width=0.95\textwidth}
\end{center}
\caption{
Schematic diagram of the optical replica synthesis through optical
modulation of electron bunch and coherent radiation in the output
undulator. Signal beam filter based on polarizer: y-polarized  light
is transmitted, while x-polarized light is reflected
}
\label{fig:dg3}

\begin{center}
\epsfig{file=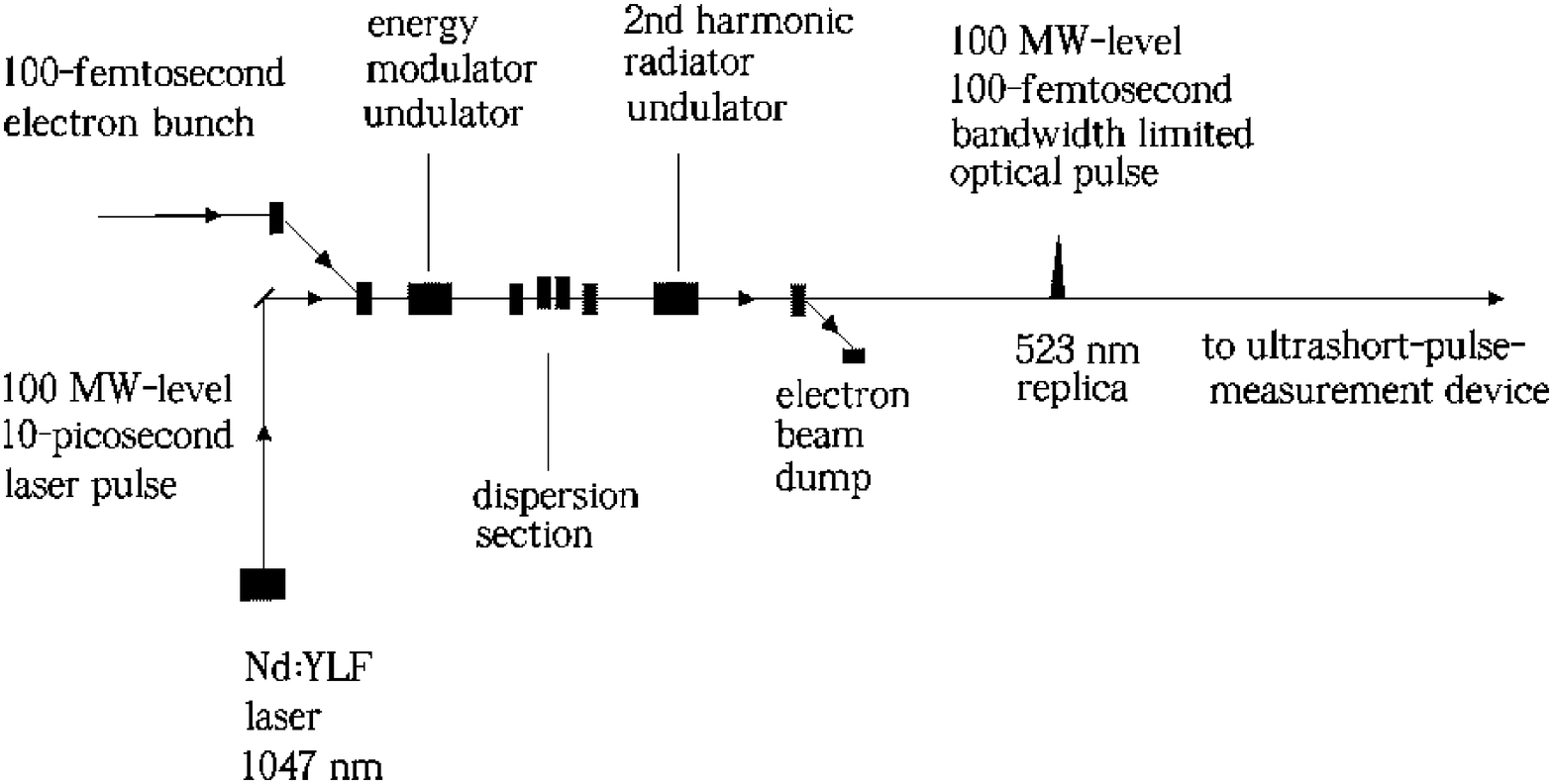,width=0.95\textwidth}
\end{center}
\caption{
Second possible schematic of replica synthesis: signal beam filter
based on the 2nd harmonic generation. The bunched beam has not only a
seed radiation frequency component, but also a considerable intensity
in its harmonics. It is then possible to have an input undulator
operating at one frequency, and an output undulator operating at double
of this frequency
}
\label{fig:dg1}
\end{figure}

\begin{figure}[tb]
\begin{center}
\epsfig{file=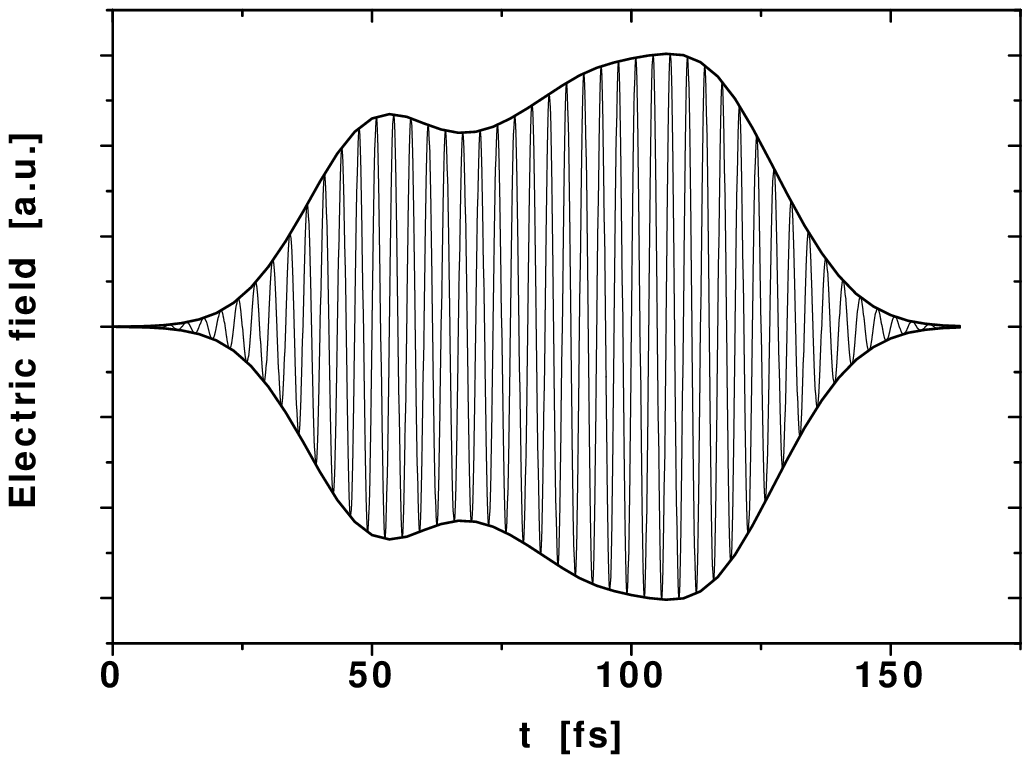,width=0.75\textwidth}
\end{center}
\caption{
Optical replica (rapidly oscillating curve) of a test electron bunch.
Radiator operates at the wavelength of 1047 nm
}
\label{fig:ex-e}
\end{figure}

A  basic scheme of the optical replica synthesizer and optical replica
of a complex test electron bunch are shown in
Figs.~\ref{fig:dg3}--\ref{fig:ex-e}. A relatively long laser pulse is
used to modulate the energy of electrons within the electron pulse at
the seed laser frequency. The electron pulse will be timed to overlap
with the central area of the laser pulse. The duration of the laser
pulse is much larger than the electron pulse time jitter of a fraction
of ps, so it can be easily synchronized with the electron pulse. The
laser pulse serves as a seed for modulator which consists of a short
undulator and dispersion section. Parameters of the seed laser are:
wavelength 1047 nm, energy in the laser pulse 1 mJ, and FWHW pulse
duration 10 ps.  The laser beam is focused onto electron beam in a
short (number of periods is equal to $N_{\mathrm{w}} = 5$) modulator
undulator resonant at the optical wavelength of 1047 nm.  Optimal
conditions of focusing correspond to the positioning of the laser beam
waist in the center of the modulator undulator. The size of the laser
beam waist is 10 times larger than the electron beam size.  The seed
laser pulse interacts with the electron beam in the modulator undulator
and produces an amplitude of the energy modulation in the electron
bunch of about 250 keV. Then the electron bunch passes through the
dispersion section (momentum compaction factor is about of $R_{56}
\simeq 50\mu$m) where the energy modulation is converted to the density
modulation at the laser wavelength. The density modulation reaches an
amplitude of about 10\%. Following the modulator the beam enters the
short (number of periods is equal to $N_{\mathrm{w}} = 5$) radiator
undulator which is resonant at laser (or double) frequency. Because the
beam has a large component of bunching, coherent emission is copiously
produces by the electron bunch. The bandwidth-limited output radiation
pulse (see Fig. \ref{fig:ex-e}) has 10 $\mu$J-level pulse energy and is
delivered in a diffraction-limited beam.

The optical replica synthesizer is expected to satisfy certain
requirements which can be achieved by suitable design and choice of the
components. A complete optimization of the proposed diagnostic device
can be performed only with three-dimensional time-dependent numerical
simulation code. Numerical results presented in this paper are obtained
with version of code FAST \cite{fast} modified for simulation of
optical replica synthesis. This code allows one to perform simulations
of coherent undulator radiation taking into account all physical
effects influencing the synthesizer operation.

\subsubsection{Low background}

One important point in the construction of replica synthesizer is
separation of the optical replica from the seed laser pulse. Numerous
designs are possible -- for example, the combination of two planar
undulators placed in crossed positions, as it is illustrated
schematically in Fig.~\ref{fig:dg3}. Both undulators have the same
period and field strength. The 1047 nm x-polarized seed radiation with
100 MW peak power and electron beam enter the first undulator, which is
used to modulate the energy of electron beam. Passing the first
undulator the beam and seed radiation enter the second undulator which
is rotated by $90^{\circ}$ relatively to the first undulator section.
The x-polarized seed radiation does not interact with the electron beam
and thus propagates freely. However, a new y-polarized radiation
component is generated by the density-modulated electron beam and
rapidly reaches 100 MW-level peak power. Then the electron and the
light beam are separated. The electron beam is guided through a bypass
and the radiation enters the polarizer which selects y-polarization.
The radiation pulse after polarizer has ultrashort duration and is
exact replica of the electron bunch. Finally, the radiation pulse is
directed to the ultrashort-pulse-measurement device.

In another scheme a frequency doubler is used to distinguish the
optical replica from the intense seed laser pulse. The bunched beam at
large values of the bunching parameter has not only a fundamental
radiation frequency component, but also a considerable intensity in its
harmonics. It is then possible to have an input undulator operating at
one frequency, and an output undulator operating at a multiple of this
frequency. The radiation in the output undulator will then be excited
by the harmonic component in the electron beam, and the diagnostic
instrument will operate as a frequency multiplier. A schematic diagram
of the 2nd harmonic replica synthesis is shown in Fig.~\ref{fig:dg1}.
Following the modulator the beam and seed radiation enter short
undulator (radiator) which is resonant with the second harmonic of the
seed radiation. In the radiator the seed radiation plays no role and is
diffracted out of the electron beam, while a new 2nd harmonic radiation
is generated by the density-modulated electron beam.

\begin{figure}[p]
\begin{center}
\epsfig{file=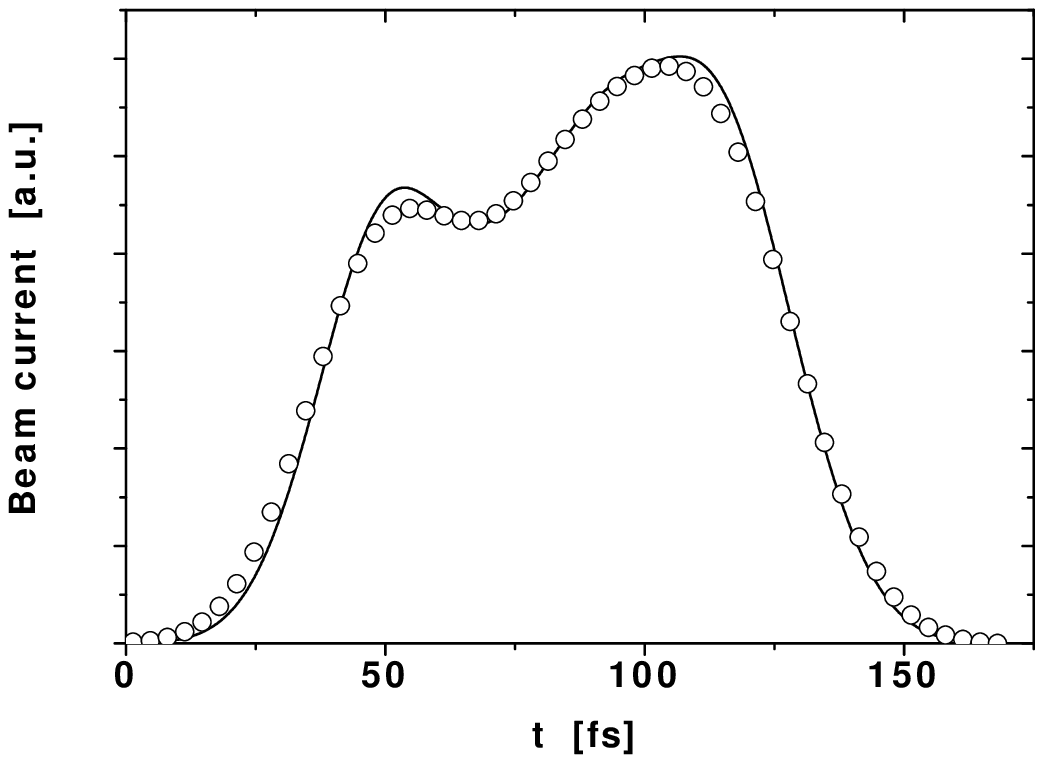,width=0.75\textwidth}
\end{center}
\caption{
Target electron beam current (solid curve) and retrieved electron pulse
shape (circles) from the optical replica in Fig. \ref{fig:ex-e}. Number
of radiator undulator periods is equal to $N_{\mathrm{w}} = 5$.  The
optical replica is generated at the radiation wavelength 1047 nm.
Discrepancies between the target and retrieved shapes are due to the
slippage effect
}
\label{fig:ex-ia}

\begin{center}
\epsfig{file=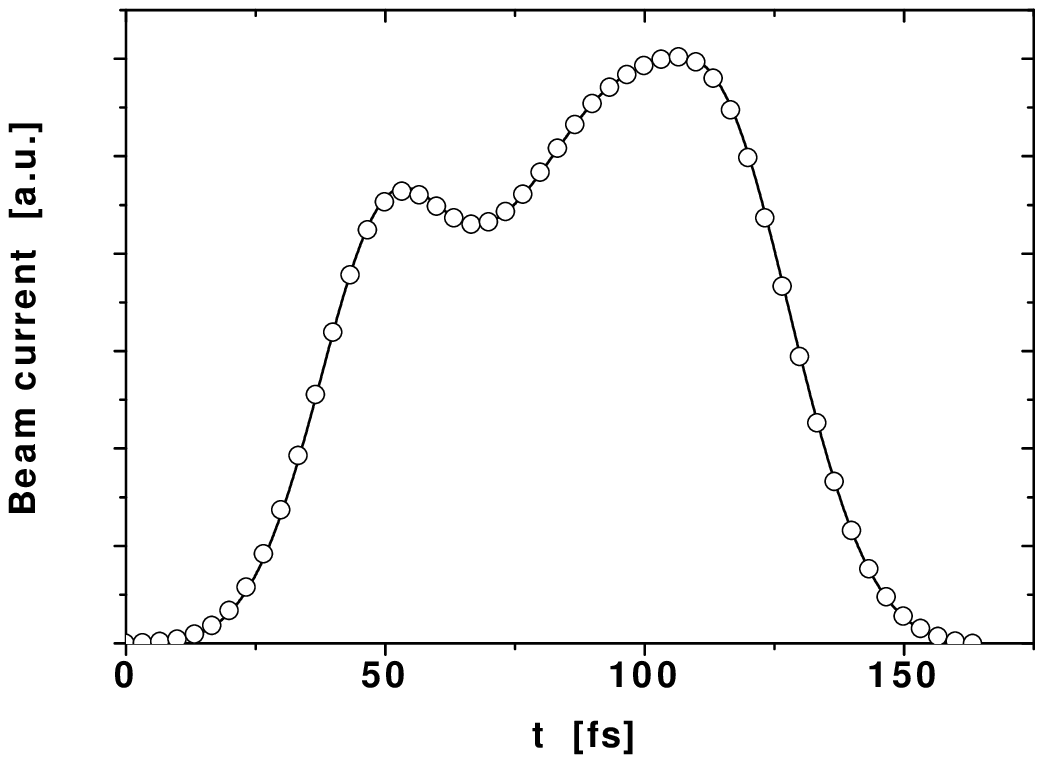,width=0.75\textwidth}
\end{center}
\caption{
Target electron beam current (solid curve) and retrieved electron pulse
shape (circles) from the optical replica. Number of radiator undulator
periods is equal to $N_{\mathrm{w}} = 5$. The optical replica is
generated at the radiation wavelength 523 nm. Note that the actual and
retrieved electron bunch shapes are visually identical
}
\label{fig:ex-ib}
\end{figure}

\subsubsection{High resolution}

When propagating in vacuum, the radiation field is faster than the
electron beam, and it moves forward (slips) by one wavelength,
$\lambda$, per one undulator period, $\lambda_{\mathrm{w}}$. It is
clear that the resolution of the electron pulse shape is determined by
the slippage of the radiation with respect to electrons in the output
undulator.  If the slippage time is much less than the electron pulse
duration,

\begin{displaymath}
N_{\mathrm{w}}\lambda/c \ll \tau_{\mathrm{e}} \ ,
\end{displaymath}

\noindent then one can neglect the slippage effect.
Calculation of the slippage effect shows (see Figs. \ref{fig:ex-ia} and
\ref{fig:ex-ib}) that this should not be a serious limitation in
our case.

\clearpage

\subsubsection{Discussion of complicating self-interaction effects}

In most applications high electron beam intensities are desired and it
is therefore prudent in particular cases to test for the appearance of
self-interaction effects. Proposed method for the electron pulse-shape
measurement is based on the assumption that beam density modulation
does not appreciably change as the beam propagates through the radiator
undulator. As any oscillating charge radiates energy, so must a
modulated electron beam moving along an undulator radiate energy. If
the system radiates energy, then in order to preserve conservation of
energy we must find that the electron beam energy is being lost.  The
electrons with different arrival phases acquire different values of the
energy increments (positive or negative), which results in the
modulation of the longitudinal velocity of the electrons with radiation
frequency. Since this velocity modulation is transformed into the
density modulation of the electron beam during the undulator pass, an
additional radiation field exists because of variation in amplitude of
density modulation.  Instead, we assume that the amplitude of electron
beam density modulation has the same value at all points in the
undulator. This approximation means that only the contributions to the
radiation field arising from the initial density modulation are taken
into account, and not those arising from the induced bunching.

The problem of induced beam density modulation in the radiator
undulator refers to the class of self-interaction problems.
Optimization of the radiator undulator length has been performed with
code FAST which takes into account collective fields (radiation and
space charge fields). Typical temporal structure of electron bunches
(mean energy, current, emittance and energy spread along the bunch) at
the exit of the bunch compression system is presented in Fig.
\ref{fig:comp}. These data sets are used as input parameters for code
FAST. The smaller the number of output undulator periods, the smaller
the induced density modulation and additionally smaller the slippage
effect. The optimum output undulator length, keeping the resonance
approximation, results in the number of periods of $N_{\mathrm{w}} =
5$. Calculation shows that in this case the ratio of the induced
density modulation amplitude and the initial amplitude at the output
undulator exit reaches value of about a few per cent only. Thus we
find that collective effects in the output undulator are not important
in our case.

Longitudinal beam dynamics in the modulator undulator as assumed in
this paper is governed by purely single-particle effects where the
results do not depend on the presence of other particles. During the
passage through a modulator the electron density modulation at the
optical wavelength can be perturbed by the collective fields.  As a
result, the small induced bunching requirement dictates the use of
modulator undulator length to be of a few periods only.  In the case
under study, the optimum number of the modulator undulator periods is
equal to $N_{\mathrm{w}} = 5$.

The next problem to be studied is that of estimating the collective
effects influencing the operation of dispersion section. Particles in a
modulated bunch following a curved path may radiate coherently at a
wavelength of the beam density modulation.  When an electron bunch
passes the dispersion section, radiative interaction induces an
additional density modulation.  The design of modulator chicane is
based on the need to minimize coherent synchrotron radiation (CSR)
induced microbunching. The problem connected with radiative interaction
of the particles in the bunch with sinusoidal density excitation moving
in a magnetic chicane has been investigated analytically and
numerically \cite{ds1,ds2,ds3}.  Calculation of the CSR effects shows
that this should not be a serious limitation in our case.

\subsubsection{The method of obtaining information about electron
current profile}

The study and detailed understanding of the cause and nature of
collective effects is important for successful design of replica
synthesizer. Proposed design is conducted to eliminate collective
effects as much as possible through installation of short input and
output undulators. The signal produced by replica synthesizer is thus a
pulse of electric field amplitude:

\begin{displaymath}
E(t) = F(I(t),\epsilon_{\mathrm{n}}(t),\Delta{\cal E}(t)) =
I(t)f(\epsilon_{\mathrm{n}}(t),\Delta{\cal E}(t)) \ ,
\end{displaymath}

\noindent where $\epsilon_{\mathrm{n}}(t)$ is the normalized slice
emittance and $\Delta{\cal E}(t)$ is the slice energy spread in the
electron bunch. If longitudinal beam dynamics in the synthesizer is
governed by purely single-particle effects then this field directly
proportional to the peak current $I(t)$.

Within the scope of the electrodynamic theory the output
characteristics of the replica synthesizer are controlled by three
dimensional parameters: $\lambda, \quad L_{\mathrm{w}}, \quad \sigma$,
where $\lambda$ is the radiation wavelength, $L_{\mathrm{w}} =
N_{\mathrm{w}}\lambda_{\mathrm{w}}$ is the radiator undulator length,
and $\sigma$ is the electron beam transverse size. At an appropriate
normalization of electrodynamic equations, the coherent undulator
radiation is described by only one dimensionless parameter:

\begin{displaymath}
N = 2\pi\sigma^{2}/(\lambda L_{\mathrm{w}}) \ .
\end{displaymath}

\noindent The parameter $N$ can be referred to as the electron beam
Fresnel number, or as diffraction parameter. In general case the
electric field of the wave radiated in the undulator depends on the
transverse size of the electron beam. For a proposed diagnostic
technique it is of great interest to minimize the influence of the
transverse emittance on the radiation field amplitude. In the case of a
wide electron beam

\begin{equation}
\lambda L_{\mathrm{w}} \ll 2\pi \sigma^{2} \ ,
\qquad {\mathrm{or}} \quad N \gg 1 \ ,
\label{eq:wide-cond}
\end{equation}

\noindent the most of the
radiation overlaps with electron beam and field of the wave is inversely
proportional to the square of electron beam

\begin{displaymath}
E(t) \propto I(t)/\sigma^{2}(t) \ .
\end{displaymath}

Reducing the particle beam cross-section by diminishing the betatron
function reduces also the size of the radiation beam and increases the
total power of output radiation. This process of reducing the beam
cross-section is, however, effective only up to some point.  Further
reduction of the particle beam size would practically no effect on the
radiation beam size and total radiation power because of diffraction
effects (see Section 4).  In the limit of a thin electron beam the
transverse radiation beam size tends to the constant value and the
dependence of the output radiation on the transverse size of the
electron beam is rather weak. The boundary between these two asymptotes
is about $\sigma^{2} \simeq \lambda L_{\mathrm{w}}$.

From the preceding discussion we may want to optimize the beam geometry
as follows. The transverse size of the electron beam has to be much
smaller then the diffraction limited radiation beam size

\begin{equation}
\sigma^{2} \ll \lambda L_{\mathrm{w}}/(2\pi) \ ,
\qquad {\mathrm{or}} \quad N \ll 1 \ ,
\label{eq:thin-cond}
\end{equation}

\noindent The radiation wavelength and the undulator length dictate the
choice of the optimum transverse size of the electron beam. Let us
present a specific numerical example. Suppose $\gamma = 10^{3}$,
$\epsilon_{\mathrm{n}} = 2\pi \mu$m, $\lambda_{\mathrm{w}} = 6.5$ cm,
$N_{\mathrm{w}} = 5$, $\lambda = 1 \mu$m. If the focusing beta function
is equal to 1 m the diffraction parameter is $N =
2\pi\sigma^{2}/(\lambda L_{\mathrm{w}}) \simeq 0.04$. We come to the
conclusion that we can treat this situation as a coherent undulator
radiation generated by a thin electron beam. This condition may be
easily satisfied in practice.

Proposed design is conducted to eliminate emittance effects as much as
possible through installation of a special electron beam focusing
system. In the radiator undulator the betatron function should reach
small values (of about 1 m) forming a narrow beam waist. The signal
generated by a replics synthesizer is thus a pulse of electric field
with amplitude:

\begin{displaymath}
E(t) = F(I(t),\epsilon_{\mathrm{n}}(t),\Delta{\cal E}(t)) =
I(t)f(\Delta{\cal E}(t)) \ .
\end{displaymath}

Optimum parameters of the dispersion section can be estimated in the
following way. The expression for the fundamental component of the
bunched beam current is $i_{1}(t) = 2I(t)J_{1}(X)$, where $X = 2\pi
R_{56}\delta{\cal E}/(\lambda {\cal E}_{0})$ is dimensionless quantity
known as the bunching parameter, $\delta{\cal E}$ is the amplitude of
energy modulation induced in the modulator undulator. The function
$J_{1}(X)$ approaches $X/2$ for small $X$; thus the microbunching
approaches $i_{1}(t) \simeq X I(t)$. We see that microbunching depends
on the choice of the dispersion section strength.  One might think that
all we have to do is to get microbunching amplitude to maximum -- we
can always increase $R_{56}$ of the dispersion section and we can
always increase output power. It is not impossible to build dispersion
section that has large $R_{56}$ function. In fact, one of the main
problems in the modulator operation is preventing the spread of
microbunching due to local energy spread in the electron beam. For
effective operation of replica synthesizer the value of suppression
factor should be close to unity.  To get a rough idea of the spread of
electron density modulation, the position of the particles within the
electron beam at the dispersion section exit has a spread which is
equal to $\Delta z^{\prime} \simeq R_{56}\Delta {\cal E}/{\cal E}_{0}$,
where $\Delta{\cal E}$ is the local energy spread in the electron
bunch. We know that uncertainty in the phase of the particles is about
$\Delta\psi \simeq 2\pi\Delta z^{\prime}/\lambda$. Therefore, a rough
estimate for the microbunching spread to be small is

\begin{equation}
2\pi R_{56}\Delta{\cal E}/{\cal E}_{0} \ll 1 \ .
\label{eq:r56-cond}
\end{equation}

\noindent The result of more careful analysis (see Section 3) shows
that in our case the optimal condition can be written as $X \simeq
0.1$, $\delta{\cal E} \simeq \max(\Delta{\cal E})/3 \simeq 250$ keV.
The amplitude of energy modulation dictates the choice of the seed
laser parameters. In our case the optimal peak power of the seed laser
is about of 100 MW.

In general, radiation field depends on the peak current, $I(t)$, local
energy spread, $\Delta{\cal E}(t)$, and local emittance,
$\epsilon_{\mathrm{n}}(t)$. However, under conditions of a thin
electron beam (\ref{eq:thin-cond}) and of a microbunching spread to be
small (\ref{eq:r56-cond}), the electric field of the wave radiated in
the replica synthesizer is directly proportional to the peak current of
the electron beam:

\begin{displaymath}
E(t) = F(I(t),\epsilon_{\mathrm{n}}(t),\Delta{\cal E}(t))
= {\mathrm{const.}}\times I(t) \ .
\end{displaymath}

\noindent Thus, conditions (\ref{eq:thin-cond}) and (\ref{eq:r56-cond})
should be treated as optimal tuning of undulator length, strength of
the dispersion section and focusing beta function for measurement of
the electron bunch profile.

\subsubsection{The method of obtaining information about slice
emittance and energy spread}

We found that longitudinal profile of the electron bunch $I(t)$ can be
reconstructed on the basis of a single-shot measurements. The next
problem is determination of slice energy spread ($\Delta{\cal E}(t)$)
and slice emittance ($\epsilon_{\mathrm{n}}(t)$). This can be done on
the basis of multishot measurements. If the electron pulse shape,
$I(t)$, is known, the local energy spread $\Delta{\cal E}(t)$ can be
determined from the dispersion section strength scan. In this way, the
problem of slice energy spread measurement is transformed into a
relatively simple task of measuring the radiation field amplitude
maximum  ($\max E(t) \propto \max i_{1}(t)$). An attempt to increase of
the amplitude of the fundamental harmonic, by increasing the strength
of dispersion section, is countered by decrease the energy spread
suppression factor. In Section 3 we demonstrate that  the microbunching
$i_{1}(t)$ has clearly a maximum

\begin{displaymath}
\max i_{1}(t) =
{\mathrm{const.}}\times\delta{\cal E}[I(t)/\Delta{\cal E}(t)] \ ,
\end{displaymath}

\noindent and the optimum strength of the dispersion section is

\begin{displaymath}
R_{56} = \frac{\lambda{\cal
E}_{0}}{2\pi\Delta{\cal E}(t)} \ .
\end{displaymath}

\noindent Thus, measuring the $\max E(t)$ is strictly equivalent to
measuring the local energy spread variations along the electron bunch:

\begin{displaymath}
I(t)/[\max{E}(t)] =
{\mathrm{const.}}\times\Delta{\cal E}(t) \ .
\end{displaymath}

\noindent Since the optimal  strength of the dispersion section is
known, that of the unknown absolute value of slice energy spread,
$\Delta{\cal E}(t)$, is easily found too.

Slice emittance can be measured in the following way. Let us consider
for illustration of the method a simple model of the electron bunch,
assuming that slice emittances are different, but Twiss parameters are
the same in all slices (more general model is discussed in section 5).
The solution in our case is to realize that in a wide electron beam
asymptote (\ref{eq:wide-cond}) the most of the radiation overlaps with
the electron beam and the field of the wave is inversely proportional
to the square of the electron bunch, $E(t) \propto I(t)/\sigma^{2}(t)$.
If the electron pulse shape, $I(t)$, is known, the problem of the slice
emittance measurement is transformed into a simple task of measuring
the radiation field amplitude in the case of a wide electron beam

\begin{displaymath}
I(t)/E(t) =
{\mathrm{const.}}\times\sigma^{2}(t) \quad {\mathrm{as}}\quad
\min(\sigma^{2}) \gg \lambda L_{\mathrm{w}}/(2\pi) \ .
\end{displaymath}

\noindent Since the value of beta function and projected emittance are
known (from a standard method using a screen and quadrupole scan), then
the unknown absolute value of slice emittance
$\epsilon_{\mathrm{n}}(t)$ is easily determined, too.

\begin{figure}[tb]

\epsfig{file=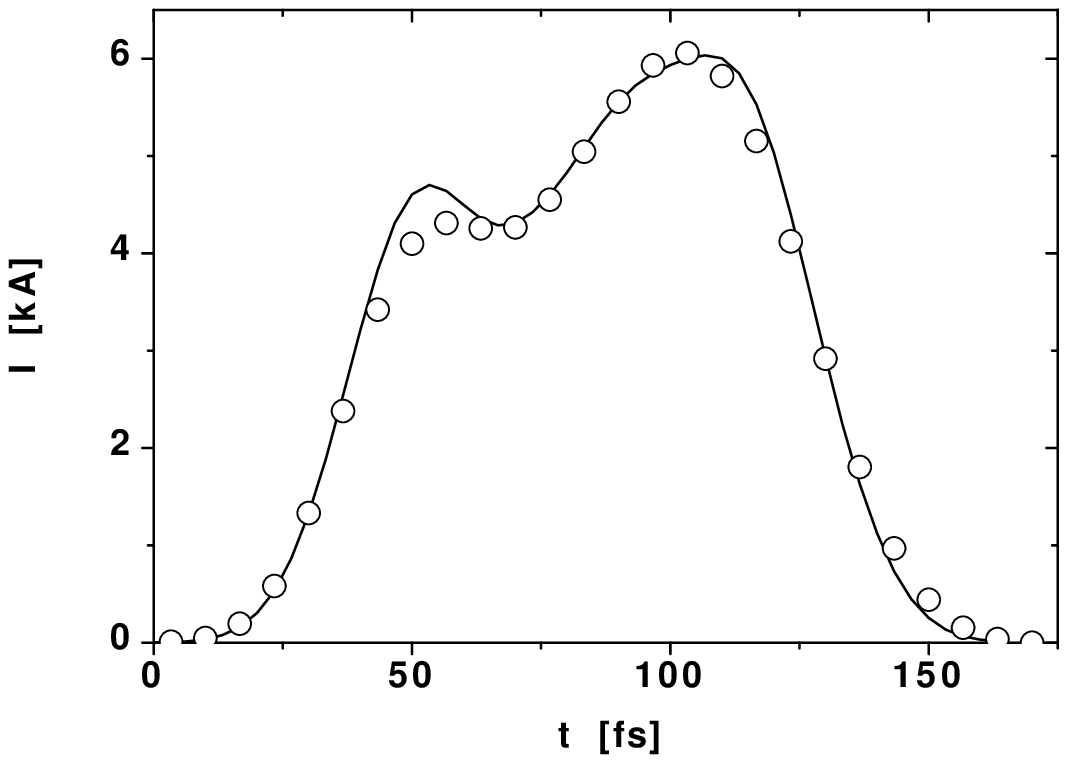,width=0.5\textwidth}

\vspace*{-60mm}

\hspace*{0.5\textwidth}
\epsfig{file=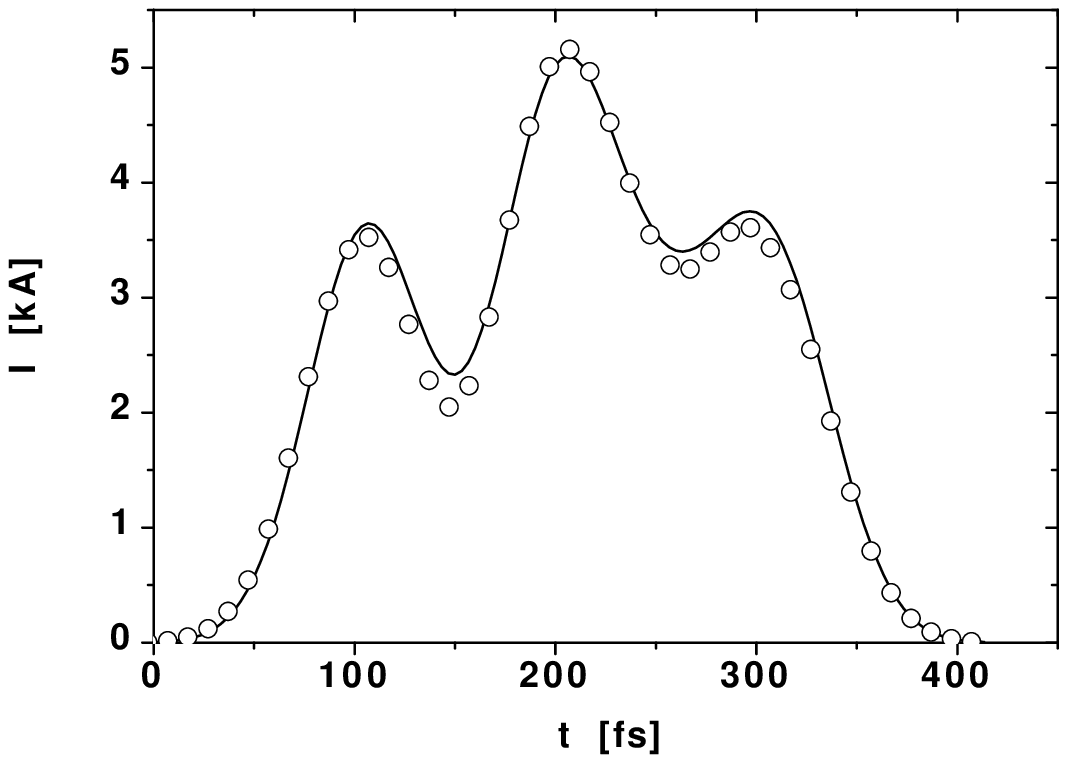,width=0.5\textwidth}

\epsfig{file=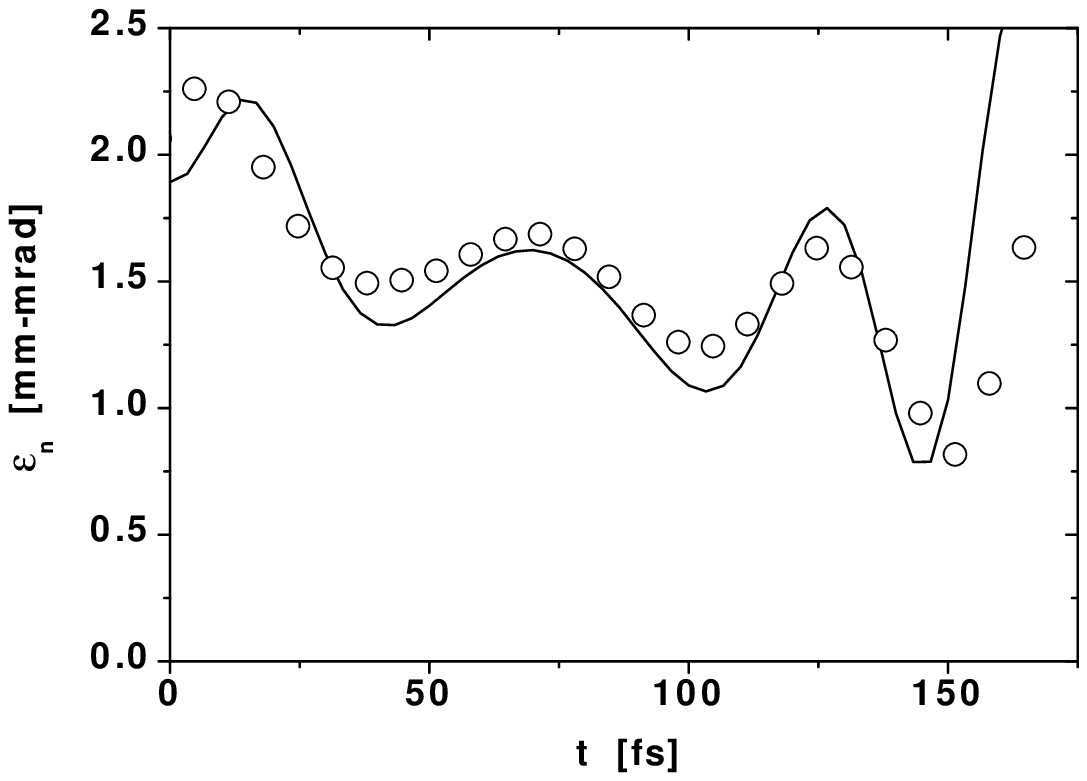,width=0.5\textwidth}

\vspace*{-60mm}

\hspace*{0.5\textwidth}
\epsfig{file=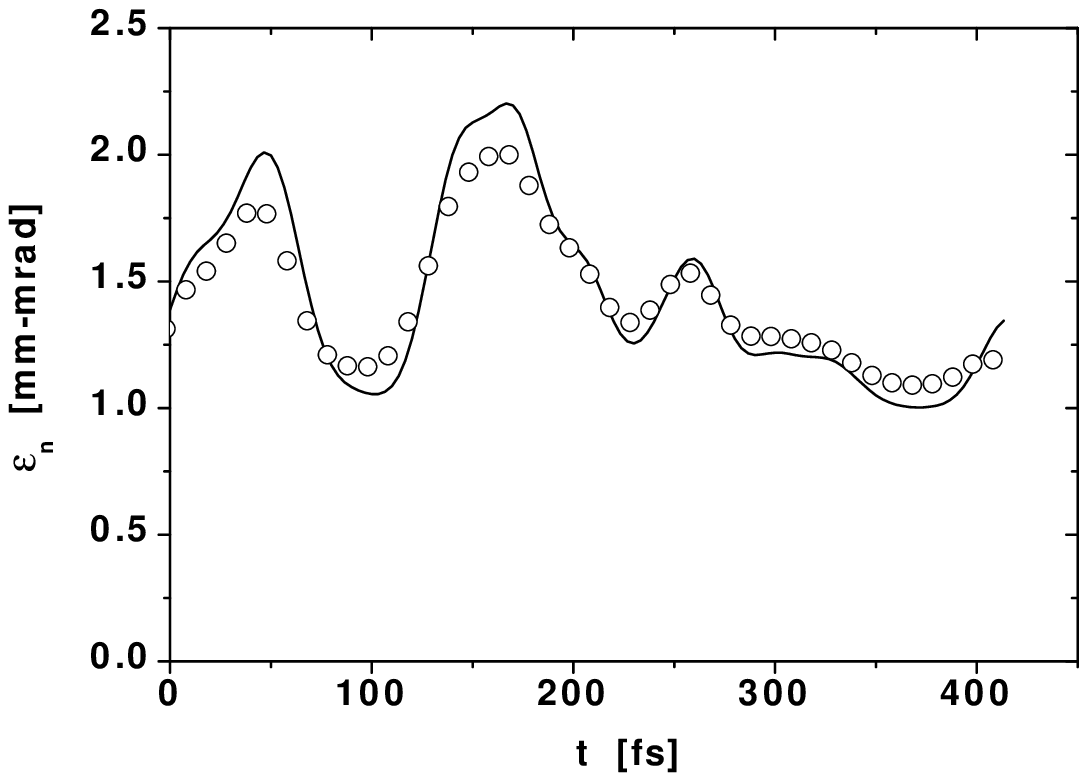,width=0.5\textwidth}

\epsfig{file=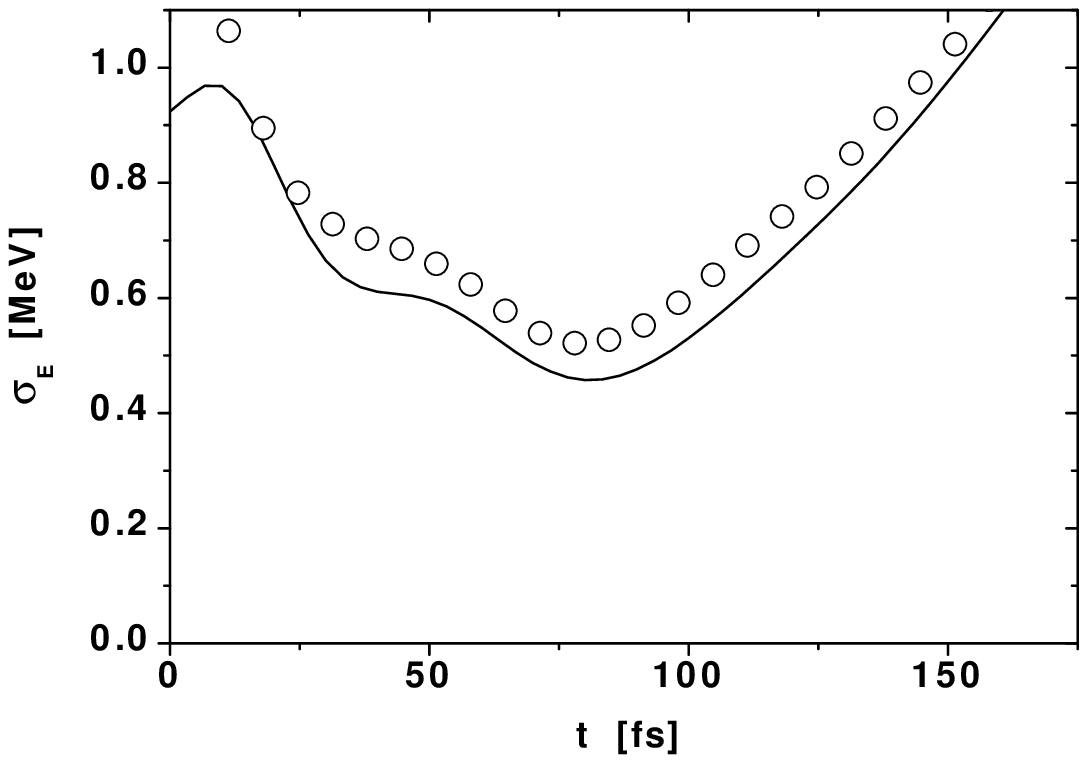,width=0.5\textwidth}

\vspace*{-60mm}

\hspace*{0.5\textwidth}
\epsfig{file=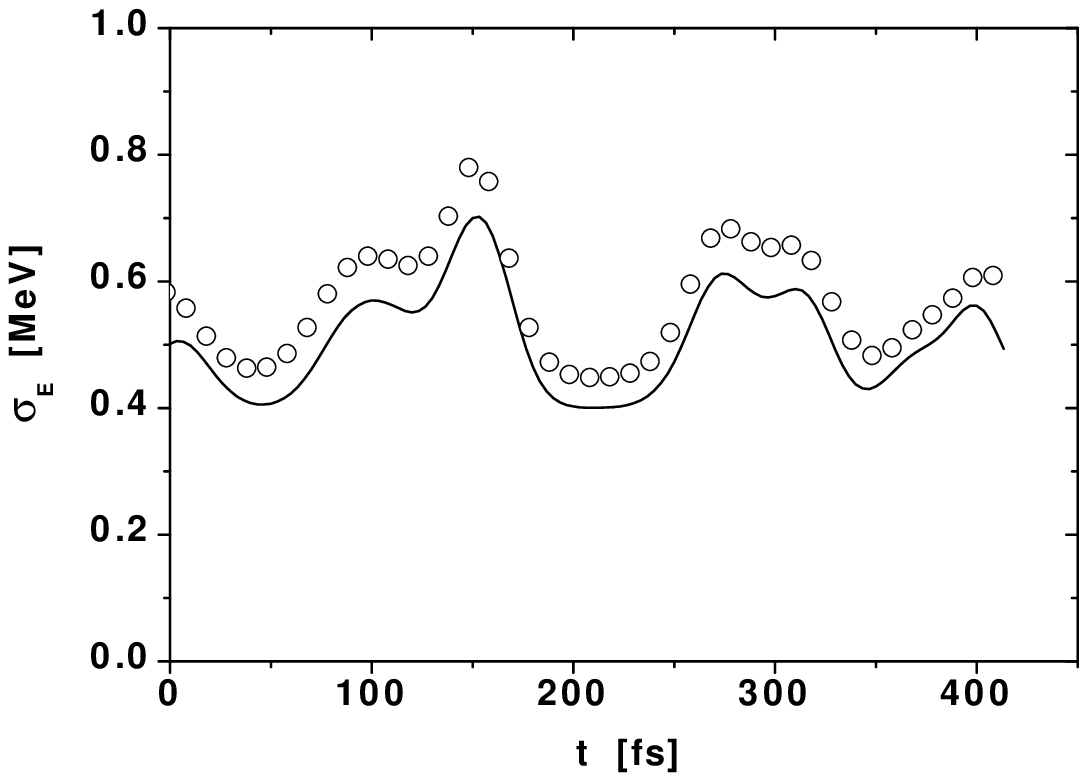,width=0.5\textwidth}

\caption{
Target (solid curve) and retrieved (circles) electron beam peak
current, slice emittance and slice energy spread.  Two different test
electron pulses are used in the optical replica calculations (right and
left columns). The nominal energy of electrons is equal to ${\cal
E}_{0} = 500$ MeV. Number of undulator periods is equal to
$N_{\mathrm{w}} = 5$. The optical replica is generated at the radiation
wavelength 1047 nm
}
\label{fig:comp}
\end{figure}

We illustrate retrieval of the slice bunch properties from the optical
replica of the electron bunch. We take two different electron bunches
(right and left columns in Fig.~\ref{fig:comp}), and perform numerical
calculations using code FAST. The nominal energy of electrons is equal
to ${\cal E}_{0} = 500$ MeV. Number of undulator periods in the
modulator and radiator undulator is equal to $N_{\mathrm{w}} = 5$.
Period length is 6.5~cm. The optical replica is generated at the
radiation wavelength 1047 nm. The seed laser power is 100~MW, FWHM
pulse duration is 10~ps. Upper plots in Fig.~\ref{fig:comp} show
comparison of target and reconstructed values for the beam current.
When taking these data, parameters for the numerical experiment were
set according to conditions (\ref{eq:thin-cond}) and
(\ref{eq:r56-cond}): focusing beta function in the radiator is 1~meter,
and net compaction factor of the dispersion section is 50~$\mu $m.
Calculations show that pulse energy in the optical replica exceeds
30~$\mu $J. Slice energy spread was determined by means of the scan of
dispersion section strength at the value of beta-function of 1~meter
(lower plots in Fig.~\ref{fig:comp}). The values of slice emittance
were extracted with the help of additional set of calculations with
large value of beat function of 50~m which corresponds to the limit of
a wide electron beam. We see that slice bunch properties can be
retrieved with high accuracy if optical replica can be characterized
with high accuracy. The latter problem is the subject of the next
section.

There is no doubt that numerical simulation code gives a correct
prediction for a given set of parameters. However, in many cases a more
transparent physical analysis would be more preferable. The calculation
scheme of the replica characteristics which is suitable for engineering
practice is presented in Sections 3 and 4. This scheme stems from
similarity techniques and numerical calculation results given as
universal plots. It may be especially useful at the design stage of an
experiment. To concentrate on the diffraction effects, in Section 4 we
have restricted our attention to the steady-state theory of the
coherent undulator radiation. We assumed that a continuous electron
beam with current density constant in time is fed to the undulator
entrance. In practical situations the electron beam has a finite pulse
duration (about 100 fs), and the question arises of when one can use
the results of Section 4. If the slippage time of the radiation with
respect to electrons per undulator length is much less than the
electron pulse duration, then one can neglect the slippage effects and
use the steady-state approach. Now let us consider the electron pulse
with the gradient axial profile of current $I(t)$.  As an
approximation, the smooth profile $I(t)$ may be replaced by a "boxcar"
function. The pulse duration interval is divided into $N_{\mathrm{s}}$
subintervals of equal length. Within each subinterval, the
approximation to $I(t)$ is constant. At the end of the each
subinterval, the approximate profile jumps to a new constant value.
When $N_{\mathrm{w}}\lambda/c < \tau_{\mathrm{e}}/N_{\mathrm{s}}$ we
can calculate the coherent undulator radiation separately within each
subinterval.  Using the plots presented in Fig. \ref{fig:ex-ia} and
Fig.  \ref{fig:ex-ib}, one can give a quantitative answer to the
question about the region of applicability of the steady-state model.

\subsection{Ultrashort optical pulse-shape measurements using
frequency-resolved optical gating}

The shape of the electron bunch replica cannot be measured using even
the fastest photodiodes or streak camera detectors. The rise time of
the best streak-cameras approaches 0.1 picosecond, far too slow to
resolve the femtosecond structure of ultrafast optical pulses. Early
on, it was realized that the only event fast enough to measure an
ultrashort pulse is the pulse itself. This gave birth to the
now-standard method of measurement: the intensity autocorrelation (AC).
Specifically, it involves splitting the pulse into two, variably
delaying one with respect to the other, and spatially overlapping the
two pulses in some instantaneously responding nonlinear-optical medium,
such as second-harmonic-generation (SHG) crystal. A SHG crystal
produces light at a twice the frequency of input light with an
intensity that is proportional to the product of the intensities of the
two input pulses. It is clear that this yields some measure of the
pulse length because no second harmonic intensity will result if the
pulses do not overlap in time. Thus, a relative delay of one pulse
length will typically reduce the SHG intensity by about a factor of
two.

\begin{figure}[tb]
\begin{center}
\epsfig{file=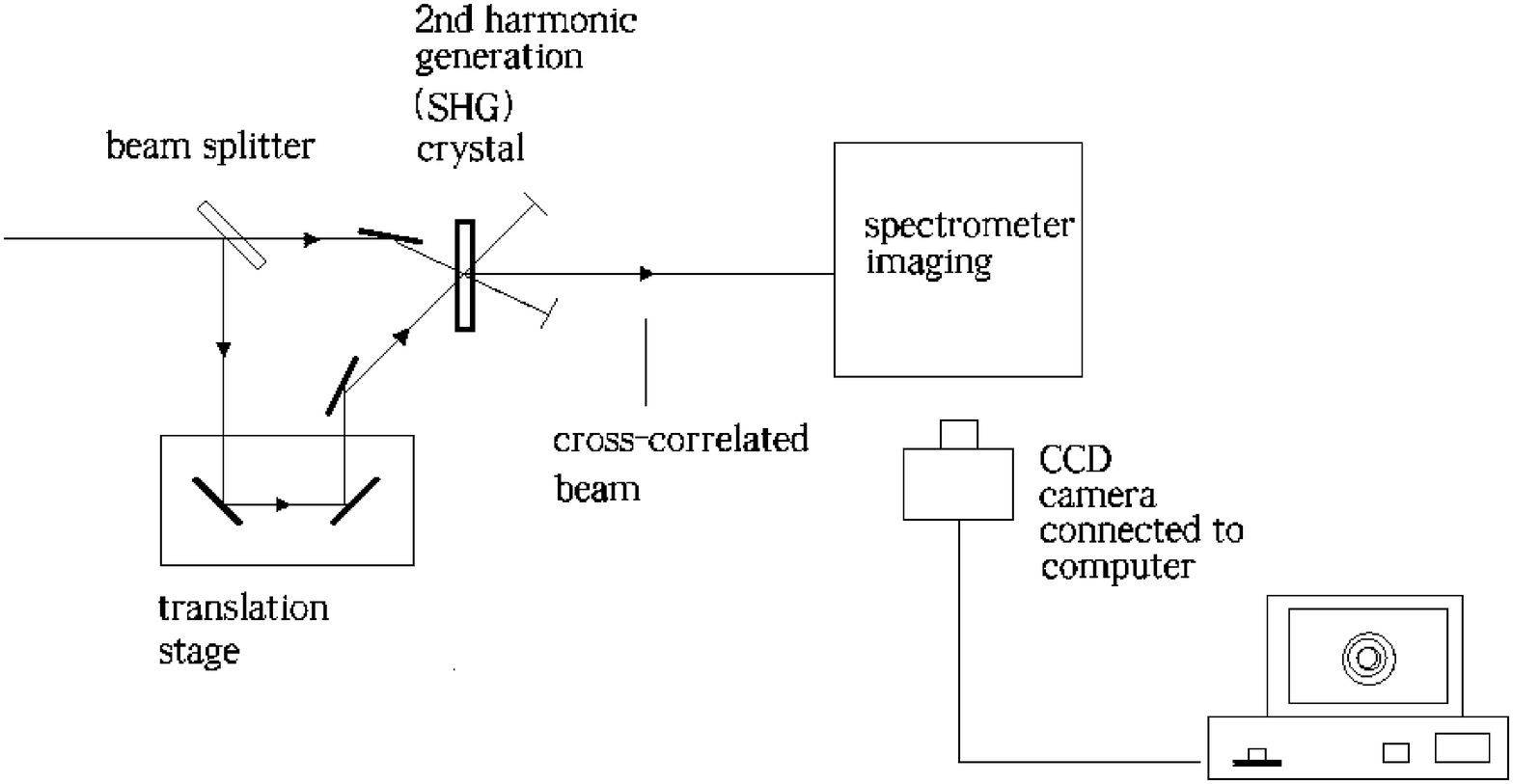,width=0.95\textwidth}
\end{center}
\caption{
A schematic of ultrashort-pulse-measurement device -- SHG FROG, the
most common and most sensitive version of FROG. Like in the
autocorrelation device the two beams are combined in the SHG crystal
and a frequency doubled signal beam is created. This signal beam is
then sent through an imaging spectrometer which outputs the beam
frequency as a function of a distance. The output of the spectrometer
(the FROG trace, or spectrogram) is captured by a CCD camera. The pulse
shape is then determined using an algorithm in a computer connected to
the camera
}
\label{fig:dg2}
\end{figure}

\begin{figure}[tb]
\begin{center}
\epsfig{file=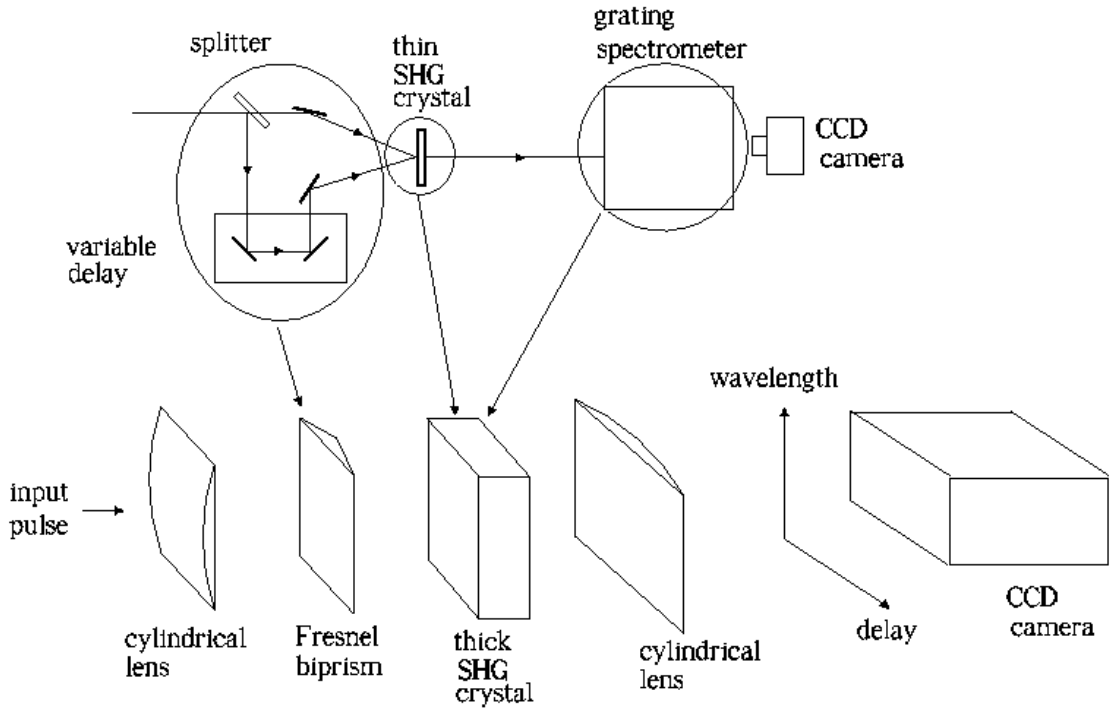,width=0.95\textwidth}
\end{center}
\caption{
A schematic of a single-shot  FROG trace measurement device. FROG trace
can be produced by a device composed of a few as five simple optical
elements. GRENOUILLE is the simplest ultrashort-pulse measurement
device in the history. This trivial device uses a Fresnel biprism to
replace the beam splitter, delay line, and beam-recombining optics. It
maps delay to position at the crystal. GRENOUILLE also utilizes a thick
SHG crystal acting as both the non-linear-optical  time-gating element
and the spectrometer. A complete single-shot SHG FROG trace results.
Whereas an autocorrelator (see top) has four sensitive alignment
parameters, GRENOUILLE has no sensitive alignment parameters at all
}
\label{fig:dg4}
\end{figure}

\begin{figure}[tb]
\begin{center}
\epsfig{file=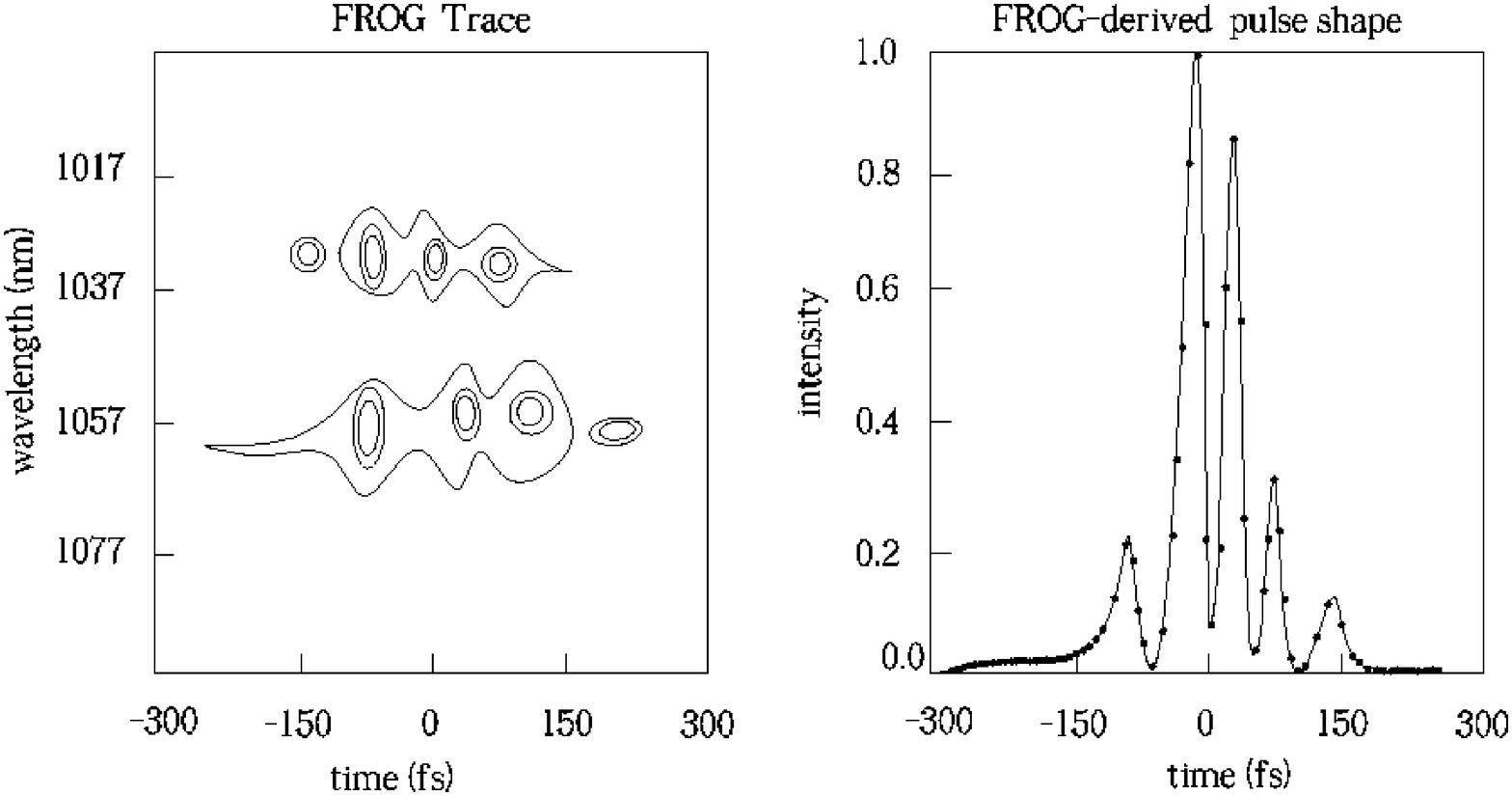,width=0.95\textwidth}
\end{center}
\caption{
Front panel display of the femtosecond oscilloscope. The device can
display the pulses inverted by an iterative algorithm at a rate of 10~Hz
}
\label{fig:dg5}
\end{figure}

Mathematically, the autocorrelation $A(\tau)$ is given by:

\begin{displaymath}
A(\tau) = \int\limits^{\infty}_{-\infty}I(t)I(t+\tau)\D t \ .
\end{displaymath}

\noindent One immediately recognizes the physical meaning of the
autocorrelation function. The Fourier transform of the autocorrelation
is $\bar{A}(\omega)$, related to the Fourier transform of the signal
by: $\bar{A}(\omega) = \mid\bar{I}(\omega)\mid^{2}$. An autocorrelation
is always a symmetric function. The Fourier transform of the
autocorrelation is a real function, consistent with a symmetric
function in the time domain. The question then naturally arises as to
exactly what information about $I(t)$ can be derived from the
measurement of cross-correlation. One can see that the correlation
technique provides the possibility to measure the modulus of the
Fourier transform of the signal function, while information about its
phase is missing. Also, even when the spectrum is also measured there
is not sufficient information to determine the pulse. Despite these
serious drawbacks, the autocorrelation and spectrum have remained the
standard measures of ultrashort pulses for over 20 years, largely for
lack of better methods \cite{ultrf}.

The problem of ultrafast-pulse measurements have recently been solved.
The autocorrelator and spectrum are the building blocks for a new
pulse-shape measurement technique, frequency resolved optical gating
(FROG), which is simply the spectrum of autocorrelation \cite{treb}. A
tandem combination of autocorrelator and spectrum can be used to
extract shape information from ultrashort pulses. The technique is
applicable to single-shot measurements. Although there are many
different types of FROG's the type of geometry we will focus on what is
known as the Second Harmonic Generation (SHG) FROG (see Fig.
\ref{fig:dg2}). Under this FROG geometry, a SHG crystal is used just
like in the autocorrelation device to resolve the time axis, but
additionally the signal beam is sent through an imaging spectrometer,
which uses diffraction gratings to separate the light of the signal
beam, in effect spatially representing the frequency of the signal
beam. The spectrometer outputs to camera the images of the signal beam
after they have been separated into its component wavelengths, known as
FROG traces. To retrieve pulse-shape information the trace is sent
through FROG algorithm, which uses constraints to iterate to a unique
solution for both the phase and intensity of the pulse as a function of
time. The intensity as a function of time will give us the structure of
the electron bunch.

Measurement of a spectrogram, that is, the Fourier transform of a
function of two variables, thus frames the ultrashort-pulse measurement
problem in a form that allows a rigorous and general solution. This
realization lead to the introduction of iterative inversion algorithms.
The important point is that any algorithm that takes into account all
the $N\times N$ data points of the spectrogram, rather than $N$ data
points in the time domain and $N$ data points in the frequency domain,
produces a better estimate of the pulse, since it has much more
materials with which to work. The problem of determining the pulse
intensity and phase from spectrogram is essentially equivalent to the
two-dimensional "phase retrieval" problem in image science. Phase
retrieval is the problem of finding a function knowing only the
magnitude (but not the phase) of its Fourier transform. Phase retrieval
for a function of one variable is impossible. For example, knowledge of
a pulse spectrum does not fully determine the pulse -- many different
pulses have the same spectrum. But image scientists found that phase
retrieval for a function of two variables is possible. Knowledge of
only the magnitude of a two-dimensional Fourier transform of a function
of two variables essentially uniquely determines the function provided
that the function is of finite extent.

Quite surprisingly, a FROG trace of a pulse can be produced by an
almost trivial device composed of as few as five simple optical
elements. This extremely simple device is called GRENOUILLE
\cite{gren}.  It involves replacing the beam splitter, delay line, and
beam combining optics with a simple element, a Fresnel biprism (a prism
with an apex angle close to $180^{\circ}$). When a Fresnel biprism is
illuminated with a wide beam, it splits the beam into two beamlets and
crossed them at an angle yielding interference fringes.  While fringes
aren't relevant to pulse measurement, crossing beams at an angle is
exactly what is required in conventional single-shot autocorrelator and
FROG beam geometries, in which the relative beam delay is mapped onto
horizontal position at the crystal (see Fig. \ref{fig:dg4}). Beams that
are split and crossed by a Fresnel biprism are automatically aligned in
space and in time, which is a significant simplification with respect
to conventional geometries. GRENOUILLE uses a thick SHG crystal, which
not only gives considerably more signal (signal strength scales as the
approximate square of the thickness), but also simultaneously replaces
the spectrometer. It operates as a single-shot device. GRENOUILLE
involves no beam-splitting, no beam-recombining, and no scanning of the
delay, and so has zero sensitive alignment degrees of freedom. Two
additional cylindrical lenses complete the device. The first
cylindrical lens must focus the beam into the thick crystal tightly
enough to yield a range of crystal incidence (and hence exit) angles
large enough to include the entire spectrum of the pulse. After the
crystal, a cylindrical lens then maps the crystal exit angle onto
position at the camera, with the wavelength as a near-linear function
of (vertical) position.

This device is capable of measuring complex pulses. This is because a
FROG traces large number of points (about 10 thousands in a 100
$\times$ 100 trace) giving sufficient information capacity to measure a
pulse with a large amount of structure. It is also possible to measure
the intensity distribution of a single ultrashort laser pulse. The
entire trace can then be obtained on a single camera image. An
iterative phase-retrieval algorithm is used to find the pulse field for
a given trace. This algorithm works well and generally converges in a
0.1 second or so at modern CPUs. The front panel of such femtosecond
oscilloscope is shown in Fig. \ref{fig:dg5}.

\section{Operation of the optical modulator}

\begin{figure}[b]
\begin{center}
\epsfig{file=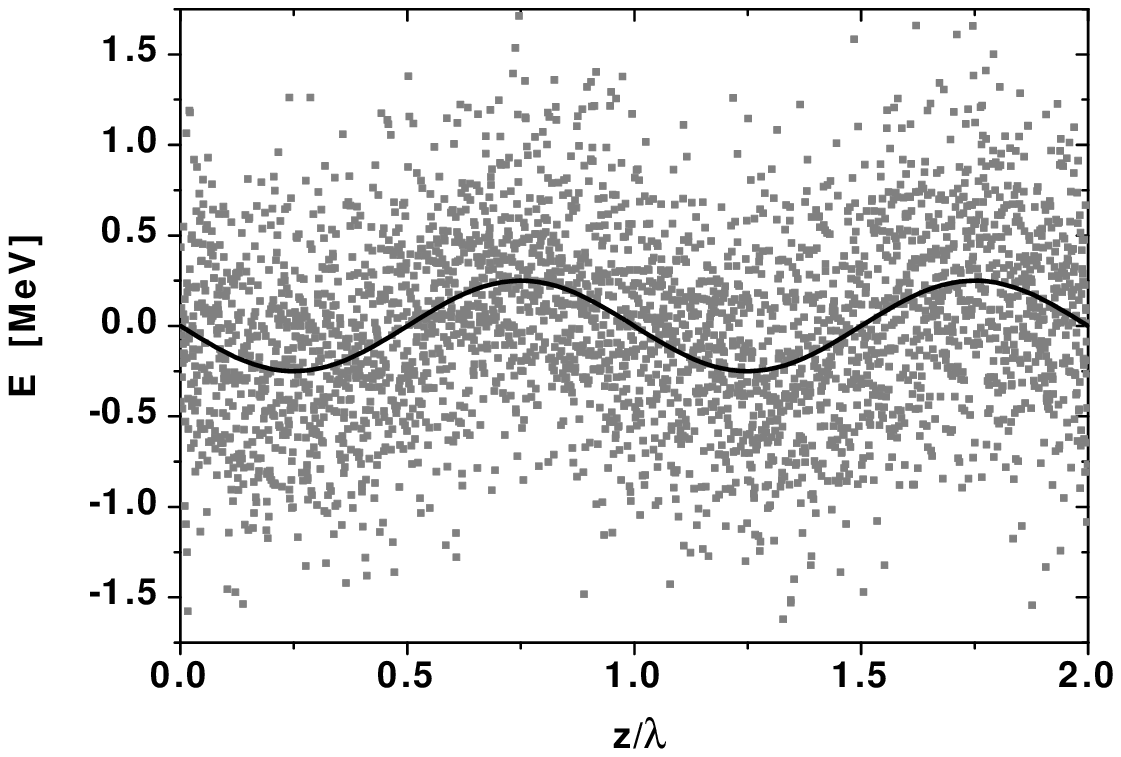,width=0.75\textwidth}
\end{center}
\caption{Phase space distribution of electrons at the exit of the
modulator undulator. Solid line shows laser induced energy modulation
}
\label{fig:enmod}

\begin{center}
\epsfig{file=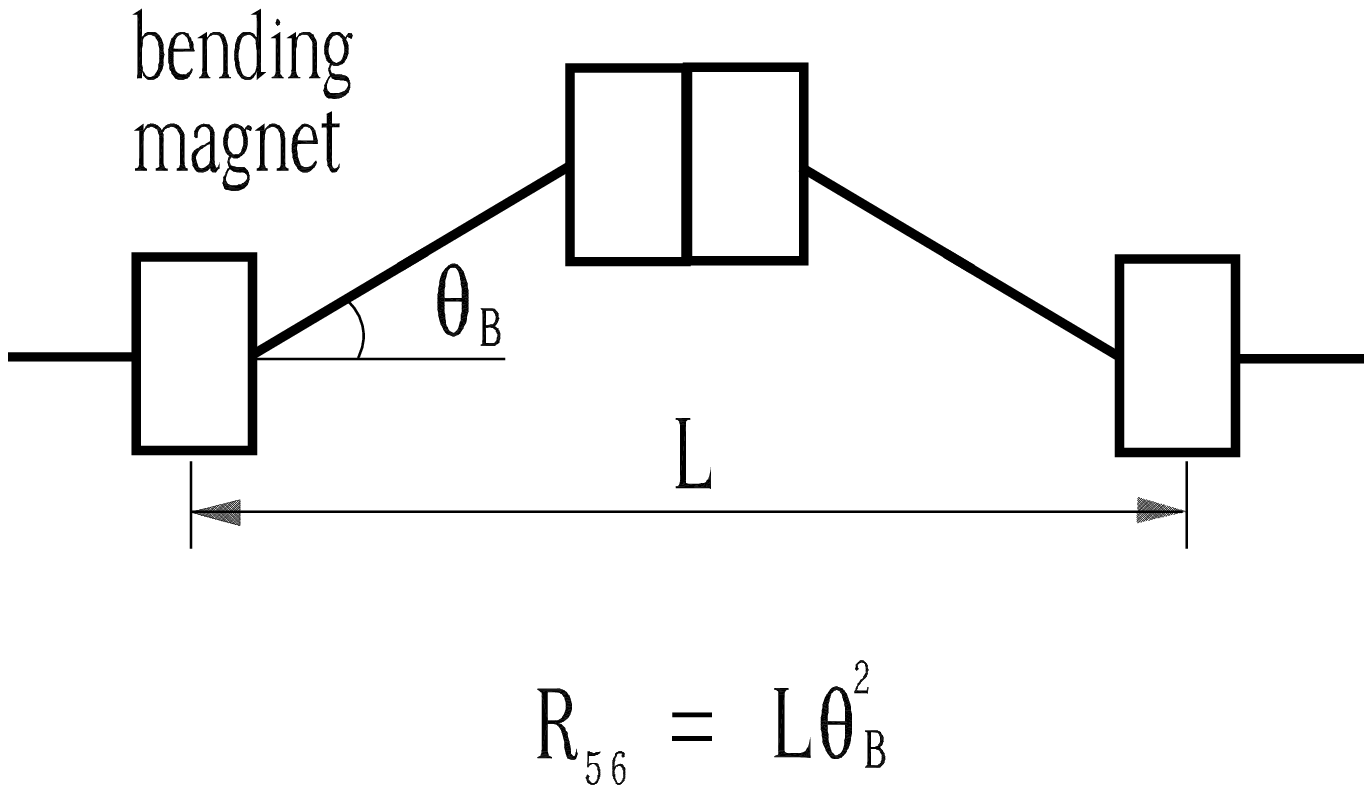,width=0.75\textwidth}
\end{center}
\caption{Schematic of dispersion section}
\label{fig:ads}
\end{figure}

The optical modulator consists of three elements: the optical seed
laser, the modulator undulator, and the dispersion section. The seed
laser pulse interacts with the electron beam in the modulator undulator
which is resonant with the laser frequency $\omega$, and produces the
energy modulation in the electron bunch (see Fig.~\ref{fig:enmod}). The
electron beam then passes through the dispersion section where the
energy modulation is converted to a density modulation at the optical
wavelength. The dispersion section is designed to introduce the energy
dependence of the particle's path length, $\Delta z = R_{56}\delta{\cal
E}/{\cal E}_{0}$. Several designs are possible, but the simplicity of a
four-dipole magnet chicane is attractive because it adds no net
beamline bend angle or offset and allows simple tuning of the momentum
compaction factor, $R_{56}$, with a single power supply (see Fig.
\ref{fig:ads}).  The trajectory of the electron beam in the chicane has
the shape of an isosceles triangle with base length $L$. The angle
adjacent to the base, $\theta_{\mathrm{B}}$, is considered to be small.
For ultra-relativistic electrons and small bend angles, the net
compaction factor $R_{56}$ of the chicane is given by

\begin{displaymath}
R_{56} = L\theta_{\mathrm{B}}^{2} \ .
\end{displaymath}

\noindent Let us consider operation of the dispersion section. The
phase space distribution of the particles in the first undulator is
described in the terms of distribution function $f(P,\psi)$ written in
"energy-phase" variables $P = {\cal E} - {\cal E}_{0}$ and $\psi = 2\pi
z^{\prime}/\lambda^{\prime} = \omega(z/v_{z}-t)$, where ${\cal E}_{0}$
is the nominal energy of the particle and $\omega$ is the angular
frequency.  Before entering the first undulator, the electron energy
distribution is assumed to be Gaussian:

\begin{displaymath}
f_{0}(P) = \frac{1}{\sqrt{2\pi\langle(\Delta{\cal E})^{2}\rangle}}
\exp\left(- \frac{P^{2}}{2\langle(\Delta{\cal E})^{2}
\rangle}\right) \ .
\end{displaymath}

\noindent The present study assumes the density modulation at the end
of the modulator undulator to be very small, and there is an energy
modulation $P_{0}\sin\psi$ only. Then the distribution function at the
entrance to the dispersion section is

\begin{displaymath}
f_{0}(P - P_{0}\sin\psi) \ .
\end{displaymath}

\noindent After passing through the dispersion section with dispersion
strength $\D\psi/\D P$, the electrons of phase $\psi$ and energy
deviation $P$ will come to the new phase $\psi + P\D\psi/\D P$. Hence
the distribution function becomes

\begin{displaymath}
f(P,\psi) = f_{0}\left(P - P_{0}\sin\left( \psi - P\frac{\D\psi}{\D
P}\right)\right)  \ .
\end{displaymath}

\noindent The dispersion strength parameter and compaction factor are
connected by the relation

\begin{displaymath}
\frac{\D\psi}{\D P} = \frac{2\pi}{\lambda^{\prime}}\frac{\D z^{\prime}}
{\D{\cal E}_{0}} = \frac{2\pi}{\lambda^{\prime}}\frac{R_{56}}{{\cal
E}_{0}} \ .
\end{displaymath}

\noindent The integration of the phase space distribution over energy
provides the beam density distribution, and the Fourier expansion of
this function gives the harmonic components of the density modulation
converted from the energy modulation \cite{czon}. At the dispersion
section exit, we may express current $I$ in the form

\begin{eqnarray}
& \mbox{} &
I =
I_{0}\int\limits^{\infty}_{-\infty}f(P,\psi)\D P = I_{0} +
2I_{0}\sum_{n=1}^{\infty}\exp\left[ - \frac{1}{2}n^{2}\langle(\Delta
{\cal E})^{2}\rangle\left(\frac{\D\psi}{\D P}\right)^{2}\right]
\nonumber\\
& \mbox{} &
\times
J_{n}\left(nP_{0}\frac{\D\psi}{\D P}\right)\cos(n\psi) \ .
\label{eq:1}
\end{eqnarray}

\noindent We find a set of harmonic waves, of which the fundamental
term, with angular frequency $\omega$, is the one of importance in a
radiator.  This fundamental term involves the phase variation
$\cos\psi$, and an amplitude term

\begin{equation}
a_{1} = 2J_{1}\left(P_{0}\frac{\D\psi}{\D P}\right)
\exp\left[ - \frac{1}{2}\langle(\Delta
{\cal E})^{2}\rangle\left(\frac{\D\psi}{\D P}\right)^{2}\right] \ .
\label{eq:a1}
\end{equation}

\noindent For small input signal we may assume that the argument of the
Bessel function is small. The function $J_{1}(X)$ approaches $X/2$ for
small $X$, thus the microbunching approaches

\begin{figure}[tb]
\begin{center}
\epsfig{file=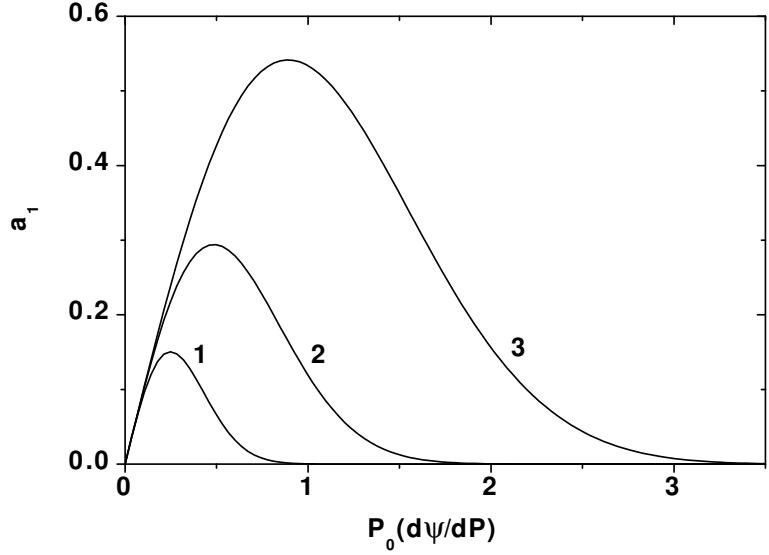,width=0.75\textwidth}
\end{center}
\caption{Dependence of $a_{1}$, fundamental component of bunched beam
current, on bunching parameter $X = P_{0}\D\psi/\D P$ and on the
parameter $P_{0}/\sqrt{\langle(\Delta{\cal E})^{2}\rangle}$ giving the
relative amplitude of beam energy modulation. Curve 1:
$P_{0}/\sqrt{\langle(\Delta{\cal E})^{2}\rangle} = 0.25$. Curve 2:
$P_{0}/\sqrt{\langle(\Delta{\cal E})^{2}\rangle} = 0.5$. Curve 3:
$P_{0}/\sqrt{\langle(\Delta{\cal E})^{2}\rangle} = 1$ }
\label{fig:dpsidp}
\end{figure}

\begin{displaymath}
a_{1} = P_{0}\frac{\D\psi}{\D P}
\exp\left[ - \frac{1}{2}\langle(\Delta
{\cal E})^{2}\rangle\left(\frac{\D\psi}{\D P}\right)^{2}\right] \ .
\end{displaymath}

\noindent The relation between $a_{1}$ and bunching parameter for
different values of energy spread is shown in Fig. \ref{fig:dpsidp}. We
see that microbunching depends greatly on the choice of the dispersion
section strength. An attempt to increase the amplitude of the
fundamental harmonic by increasing the strength of dispersion section,
is countered by a decrease of the exponential factor. The microbunching
$a_{1}$ has clearly a maximum

\begin{displaymath}
(a_{1})_{\mathrm{max}} = \frac{P_{0}}{\sqrt{2.72
\langle(\Delta{\cal E})^{2}\rangle}} \ ,
\end{displaymath}

\noindent and the optimum strength of the dispersion section is

\begin{displaymath}
\left(\frac{\D\psi}{\D P}\right)_{\mathrm{max}} = \frac{1}
{\sqrt{\langle(\Delta{\cal E})^{2}\rangle}} \ .
\end{displaymath}

\noindent Let us consider numerical example for $P_{0} = 250$ keV,
$\sqrt{\langle (\Delta{\cal E})^{2}\rangle} = 500$ keV, ${\cal E}_{0} =
500$ MeV, and $\lambda = 1047$ nm. The calculation gives $a_{1} = 0.1$
at $R_{56} = 30 \mu$m. The suppression factor in (\ref{eq:a1}) is equal
to $\exp[- \omega^{2}\langle(\Delta{\cal E})^{2}\rangle R^{2}_{56}
/(2c^{2}{\cal E}^{2}_{0})] \simeq 0.98$. We come to the conclusion that
we can treat this situation as an optimum modulator design. The optimum
condition for the replica synthesis is that the modulator should
present  a rather weak dependence of the output modulation on the local
energy spread in the electron beam.

\section{Operation of the optical radiator}

Proposed technique of electron bunch diagnostics essentially exploits
the properties of the radiation generated by modulated electron beam in
the undulator. To simplify consideration we start with the case of a
helical undulator. Later on all the results will be generalized for the
case of a planar undulator. The magnetic field on the axis of the
helical undulator is given by

\begin{displaymath}
\vec{H}_{\mathrm{w}} = \vec{e}_{x}H_{\mathrm{w}}\cos(k_{\mathrm{w}}z)
- \vec{e}_{y}H_{\mathrm{w}}\sin(k_{\mathrm{w}}z) \ ,
\end{displaymath}

\noindent where $k_{\mathrm{w}} = 2\pi/\lambda_{\mathrm{w}}$ is the
undulator wavenumber and $\vec{e}_{x,y}$ are unit vectors directed
along the $x$ and $y$ axes of the Cartesian coordinate system
$(x,y,z)$. The Lorentz force $\vec{F} =
-e(\vec{v}\times\vec{H}_{\mathrm{w}})/c$ is used to derive the
equations of motion of electrons with charge $(-e)$ and mass
$m_{\mathrm{e}}$ in the presence of the magnetic field. The explicit
expression for the electron velocity in the field of the helical
undulator has the form:

\begin{displaymath}
\vec{v}_{\perp}(z) =
c\theta_{\mathrm{w}}[\vec{e}_{x}\cos(k_{\mathrm{w}}z)
- \vec{e}_{y}\sin(k_{\mathrm{w}}z)] \ ,
\end{displaymath}

\noindent which means that the electron in the undulator moves along a
constrained helical trajectory parallel to the $z$ axis. The angle of
rotation is given by the relation $\theta_{\mathrm{w}} = K/\gamma =
\lambda_{\mathrm{w}}eH_{\mathrm{w}}/(2\pi m_{\mathrm{e}}c^{2}\gamma)$,
where $\gamma = (1 - v^{2}/c^{2})^{-1/2}$ is the relativistic factor
and $v^{2} = v_{x}^{2} + v_{y}^{2} + v_{z}^{2}$. As a rule, the
electron rotation angle $\theta_{\mathrm{w}}$ is small and the
longitudinal electron velocity $v_{z}$ is close to the velocity of
light, $v_{z} \simeq c$.

Let us consider periodically modulated relativistic electron beam
moving along the $z$ axis in the field of a helical undulator. In what
follows we use the following assumptions: i) electrons move along
constrained helical trajectories in parallel with the $z$ axis; ii) the
radius of the electron rotation in the undulator, $r_{\mathrm{w}} =
\theta_{\mathrm{w}}/k_{\mathrm{w}}$, is much less than the transverse
size of the electron beam; iii) electron beam density at the undulator
entrance is simply $n = n_{0}(\vec{r}_{\perp}) [1 +
a_{\mathrm{in}}\cos\omega(z/v_{z}-t)]$, where $a_{\mathrm{in}} =$
const. In other words we consider the case in which there are no
variations in amplitude and phase of the density modulation in the
transverse plane. Under this assumptions the transverse current density
may be written in the form

\begin{displaymath}
\vec{j}_{\perp} = -e\vec{v}_{\perp}n(\vec{r}_{\perp},z/v_{z}
-t) = -e\vec{v}_{\perp}n_{0}(\vec{r}_{\perp})
[1 + a_{\mathrm{in}}\cos\omega(z/v_{z}-t)] \ ,
\end{displaymath}

\noindent where we calibrated the time in such a way that current
density has its maximum at time  $t = 0$, at point $z = 0$. Even though
the measured quantities are real, it is more convenient to use complex
representation. For this reason, starting with real $\vec{j}_{\perp}$,
one defines the complex transverse current density:

\begin{equation}
j_{x} + \I j_{y} = -ec\theta_{\mathrm{w}}n_{0}(\vec{r}_{\perp})
\exp( - \I k_{\mathrm{w}}z)
[1 + a_{\mathrm{in}}\cos\omega(z/v_{z}-t)] \ .
\label{eq:cu}
\end{equation}

\noindent Transverse current have the angular frequency $\omega$ and
two waves travelling in the same direction with variations $\exp\I
(\omega z/v_{z}-k_{\mathrm{w}}z-\omega t)$ and $\exp-\I(\omega z/v_{z}
+k_{\mathrm{w}}z-\omega t)$ will add to give a total current
proportional to $\exp(-\I k_{\mathrm{w}}z)\cos\omega(z/v_{z}-t)$. The
factor $\exp\I(\omega z/v_{z} - k_{\mathrm{w}}z-\omega t)$ indicates a
fast wave, while the factor  $\exp\I(\omega z/v_{z} +
k_{\mathrm{w}}z-\omega t)$ indicates a slow wave. The use of the word
"fast" ("slow") here implies a wave with phase velocity faster (slower)
than the beam velocity.

Now we should consider the electrodynamic problem. Using Maxwell's
equations, we can write the equation for the electric field

\begin{displaymath}
c^{2}\vec{\nabla}\times(\vec{\nabla}\times\vec{E}) = -
\partial^{2}\vec{E}/\partial t^{2} - 4\pi\partial\vec{j}/\partial t \ .
\end{displaymath}

\noindent Then we make use of the identity

\begin{displaymath}
\vec{\nabla}\times(\vec{\nabla}\times\vec{E}) = \vec{\nabla}
(\vec{\nabla}\cdot\vec{E}) - \vec{\nabla}^{2}\vec{E} \ ,
\end{displaymath}

\noindent where $\vec{\nabla}\cdot\vec{E}$ can be found from the
Poisson equation. Finally, we come to the inhomogeneous wave equation
for $\vec{E}$:

\begin{equation}
c^{2}\vec{\nabla}^{2}\vec{E} - \partial^{2}\vec{E}/\partial t^{2}
= 4\pi c^{2}\vec{\nabla}\rho + 4\pi\partial\vec{j}/\partial t \ .
\label{eq:a1a}
\end{equation}

\noindent This equation allows one to calculate the electric field
$\vec{E}(\vec{r},t)$ for given charge and current sources,
$\rho(\vec{r},t)$ and $\vec{j}(\vec{r},t)$. Thus, equation
(\ref{eq:a1a}) is the complete and correct formula for radiation.
However, we want to apply it to a still simpler circumstance in which
the second term (or, the current term) in the  right-hand side of
(\ref{eq:a1a}) provides the main contribution to the value of the
radiation field. Since in the paraxial approximation the radiation
field has only transverse components, we are interested in the
transverse component of (\ref{eq:a1a}).  Thus we consider the wave
equation

\begin{equation}
c^{2}\vec{\nabla}^{2}\vec{E}_{\perp} -
\partial^{2}\vec{E}_{\perp}/\partial t^{2} =
4\pi\partial\vec{j}_{\perp}/\partial t \ ,
\label{eq:e1}
\end{equation}

\noindent which relates the transverse component of the electric field
to the transverse component of current density.

We wish to examine the case when the phase velocity of the current wave
is close to the velocity of light. This requirement may be met under
resonance condition

\begin{equation}
\omega/c = \omega/v_{z} - k_{\mathrm{w}} \ .
\label{eq:a2}
\end{equation}

\noindent First we may point out that the statement of (\ref{eq:a2}),
the condition for the relation between $\omega, k_{\mathrm{w}}$ and
$v_{z}$, is the condition for synchronism between the transverse
electromagnetic wave and the fast transverse current wave with the
propagating constant $\omega/v_{z} - k_{\mathrm{w}}$. With a current
wave travelling with the same phase speed as the electromagnetic wave,
we have a possibility of (space) resonance between electromagnetic wave
and electrons. If this is the case, cumulative interaction between
bunched electron beam and transverse electromagnetic wave takes place.
We are therefore justified in considering the contributions of all the
waves except the synchronous one to be negligible.

Any state of transverse electromagnetic wave can always be written
as a linear combination of the two base states (polarizations). By
given the amplitudes and phases of these base states we completely
describe the electromagnetic wave state. It is usually best to start
with the form which is physically clearest. We choose the Cartesian
base states and seek the solution for $\vec{E}_{\perp}$ in the form

\begin{equation}
E_{x,y} = \tilde{E}_{x,y}(z,\vec{r}_{\perp})\exp[\I\omega(z/c-t)] +
{\mathrm{C.C.}}
\label{eq:a3}
\end{equation}

\noindent Here and in what follows, complex amplitudes related to the
field are written with a tilde. The description of the field given by
(\ref{eq:a3}) is quite general. However, the usefulness of the concept
of carrier wave number is limited to the case where the amplitude is
slowly varying function of $z$.

To determine the form of $\tilde{E}_{x,y}(z,\vec{r}_{\perp})$ we
substitute (\ref{eq:cu}) and (\ref{eq:a3}) into (\ref{eq:e1}), and have
inside the undulator

\begin{eqnarray}
& \mbox{} &
\exp[\I\omega(z/c-t)]\left\{\vec{\nabla}^{2}_{\perp} +
\frac{2\I\omega}{c}\frac{\partial}{\partial z} +
\frac{\partial^{2}}{\partial z^{2}}\right\}{\tilde{E}_{x} \choose
\tilde{E}_{y} } + {\mathrm{C.C.}} \nonumber\\ & \mbox{} & =
4\pi\frac{\omega}{c}{\cos(k_{\mathrm{w}}z)\choose
-\sin(k_{\mathrm{z}}z)}e\theta_{\mathrm{w}}
a_{\mathrm{in}}n_{0}(\vec{r}_{\perp})\sin\omega(z/v_{z}-t) \ .
\label{eq:2}
\end{eqnarray}

\noindent Here $\vec{\nabla}^{2}_{\perp}$ is the Laplace operator in
transverse coordinates. At $z > L_{\mathrm{w}}$ the right-hand side
of (\ref{eq:2}) is equal to zero.

Now we have apparently simple pair of equations -- and they are still
exact. We simplify the equations by noting that for a radiation field
it is reasonable to assume that $\tilde{E}_{x,y}$ are slowly varying
function of $z$ so that $\partial^{2}\tilde{E}_{x,y}/\partial z^{2}$
may be neglected. The corresponding requirement for the complex
amplitude is $\mid\partial\tilde{E}_{x,y}/\partial z\mid \ll
k\mid\tilde{E}_{x,y}\mid$. In other words, the radiation pulse must not
change significantly while travelling through a distance comparable
with the wavelength $\lambda = 2\pi/k$. This assumption is not a
restriction. Such is the case in all practical cases of interest.
Differential equations becomes

\begin{eqnarray}
& \mbox{} &
\exp[\I\omega(z/c-t)]\left\{\vec{\nabla}^{2}_{\perp} +
\frac{2\I\omega}{c}\frac{\partial}{\partial z}
\right\}{\tilde{E}_{x}
\choose \tilde{E}_{y} } + {\mathrm{C.C.}} \nonumber\\ & \mbox{} & =
4\pi\frac{\omega}{c}{\cos(k_{\mathrm{w}}z)\choose
-\sin(k_{\mathrm{z}}z)}e\theta_{\mathrm{w}}
a_{\mathrm{in}}n_{0}(\vec{r}_{\perp})\sin\omega(z/v_{z}-t) \ .
\label{eq:e2}
\end{eqnarray}

\noindent Although equations (\ref{eq:e2}) cannot be solved in general,
we will solve them for some special cases. These equations can be
simplified further by noting that the complex amplitudes
$\tilde{E}_{x,y}$ will not vary much with $z$, especially in comparison
with the oscillating terms $\exp(-\I k_{\mathrm{w}}z)$. The slow wave
of transverse current oscillates very rapidly about an average value of
zero and, therefore, does not contribute very much to the rate of
change of $\tilde{E}_{x,y}$. So we can make a reasonably good
approximation by replacing these terms by their average value, namely
zero. We will leave them out, and take as our approximation:

\begin{equation}
\vec{\nabla}^{2}_{\perp}{\tilde{E}_{x}
\choose \tilde{E}_{y} }
+ \frac{2\I\omega}{c}\frac{\partial}{\partial z}{\tilde{E}_{x}
\choose \tilde{E}_{y} }
=-{i \choose 1} 2\pi\frac{\omega}{c}e\theta_{\mathrm{w}}
a_{\mathrm{in}}n_{0}(\vec{r}_{\perp})\exp(-\I Cz) \ .
\label{eq:a4}
\end{equation}

\noindent Even the remaining terms, with exponents proportional to $C =
\omega/v_{z}-\omega/c-k_{\mathrm{w}}$ will also vary rapidly unless $C$
is near zero. Only then will the right-hand side vary slowly enough
that any appreciable amount will accumulate when we integrate the
equations with respect to $z$. The required conditions will be met if
$C \ll k_{\mathrm{w}} \ , \quad 1 \ll k_{\mathrm{w}}z$. In other words,
we use the resonance approximation here and assume that complex
amplitudes $\tilde{E}_{x,y}$ are slowly varying in the longitudinal
coordinate. By "slowly varying" we mean that
$\mid\partial\tilde{E}_{x,y}/\partial z\mid \ll
k_{\mathrm{w}}\mid\tilde{E}_{x,y}\mid$. For this inequality to be
satisfied, the spatial variation of $\tilde{E}_{x,y}$ within the
undulator period $\lambda_{\mathrm{w}} = 2\pi/k_{\mathrm{w}}$ has to be
small.

Equations (\ref{eq:a4}) are simple enough and can be solved in any
number of ways.  One convenient way is the following. Taking the sum
and the difference of the two we get

\begin{equation}
\left(\vec{\nabla}^{2}_{\perp}
+ \frac{2\I\omega}{c}\frac{\partial}{\partial z}\right)(\tilde{E}_{x} +
\I\tilde{E}_{y})
= 2\pi\I\frac{\omega}{c}e\theta_{\mathrm{w}}
a_{\mathrm{in}}n_{0}(\vec{r}_{\perp})\exp(-\I Cz) \ ,
\label{eq:a5}
\end{equation}

\begin{equation}
\left(\vec{\nabla}^{2}_{\perp}
+ \frac{2\I\omega}{c}\frac{\partial}{\partial z}\right)(\tilde{E}_{x} -
\I\tilde{E}_{y})
= 0 \ .
\label{eq:a6}
\end{equation}

\noindent These equations describe the general case of electromagnetic
wave radiation by the modulated electron beam in the helical undulator.
Equations (\ref{eq:a5}) and (\ref{eq:a6}) refer to the right- and
left-helicity components of the wave, respectively. The solutions for
the right- and left-helicity waves are linearly independent. It follows
from (\ref{eq:a5}) and (\ref{eq:a6}) that only those waves are radiated
that have the same helicity as undulator field itself.

The electric field, $\vec{E}_{\perp}$, of the wave radiated in the
helical undulator in resonance approximation is circularly polarized and
may be represent in the complex form:

\begin{displaymath}
E_{x} + \I E_{y} =
\tilde{E}(z,\vec{r}_{\perp})\exp[\I\omega(z/c-t)] \ .
\end{displaymath}

\noindent Finally, the equation for $\tilde{E}$ can be written in the
form

\begin{equation}
\left(\vec{\nabla}^{2}_{\perp}
+ \frac{2\I\omega}{c}\frac{\partial}{\partial z}\right)\tilde{E}
= 2\pi\I\frac{\omega}{c}e\theta_{\mathrm{w}}
a_{\mathrm{in}}n_{0}(\vec{r}_{\perp})\exp(-\I Cz) \ .
\label{eq:a7}
\end{equation}

\noindent Equation (\ref{eq:a7}) is an inhomogeneous parabolic
equation.  Its solution can be expressed in terms of a convolution of
the free-space Green's function (impulse response)

\begin{displaymath}
G(z-z^{\prime},\vec{r}_{\perp}-\vec{r}^{\prime}_{\perp}) =
\frac{1}{4\pi(z-z^{\prime})}\exp\left[\frac{\I\omega\mid\vec{r}_{\perp}
-\vec{r}^{\prime}_{\perp}\mid^{2}}{2c(z-z^{\prime})}\right]
\end{displaymath}

\noindent with the source term. When the right-hand side of
(\ref{eq:a7}) is equal to zero, it transforms to the well-known
paraxial wave equation in optics.

The radiation process displays resonance behavior, and the amplitude of
electric field depends strongly on the value of the detuning parameter
$C$. With the approximation made in getting (\ref{eq:a7}) the equation
can be solved exactly. Now we will find an exact solution for the case
of perfect resonance. When the parameters are tuned to perfect
resonance, such that $C = 0$, the solution of the equation
(\ref{eq:a7}) has the form

\begin{equation}
\tilde{E}(z,\vec{r}_{\perp}) = \frac{\I
e\theta_{\mathrm{w}}\omega a_{\mathrm{in}}}
{2c}\int\limits^{z}_{0}\frac{\D
z^{\prime}}{z-z^{\prime}}\int\D\vec{r}_{\perp}n_{0}
(\vec{r}_{\perp}^{\prime})
\exp\left[\frac{\I\omega\mid\vec{r}_{\perp}
-\vec{r}^{\prime}_{\perp}\mid^{2}}{2c(z-z^{\prime})}\right] \ ,
\label{eq:a8}
\end{equation}

\noindent where $(z,\vec{r}_{\perp})$ and
$(z^{\prime},\vec{r}^{\prime}_{\perp})$ are the coordinates of the
observation and the source point, respectively.

Let us consider an axisymmetric electron beam with gradient profile of
the current density. In this case we have
$-ev_{z}n_{0}(\vec{r}_{\perp}) = - j_{0}S(r)$, where $r$ is the radial
coordinate of the cylindrical system $(r,\phi,z)$ and $S(r)$ describes
the transverse profile of the electron beam. To be specific, we write
down all the expressions for the case of a Gaussian transverse
distribution:

\begin{displaymath}
-ev_{z}n_{0}(\vec{r}_{\perp}) = - \frac{I_{0}}{2\pi\sigma^{2}}
\exp\left(-\frac{r^{2}}{2\sigma^{2}}\right) \ ,
\end{displaymath}

\noindent where $I_{0}$ is the total beam current. Then
we can write (\ref{eq:a8}) in the form

\begin{eqnarray}
& \mbox{} &
\tilde{E}(z,r) = \frac{\I
e\theta_{\mathrm{w}}\omega a_{\mathrm{in}}I_{0}}
{2c}\int\limits^{z}_{0}\frac{\D
z^{\prime}}{z-z^{\prime}}
\int\limits^{\infty}_{0}\D r^{\prime}
r^{\prime}\exp\left[-\frac{(r^{\prime})^{2}}{2\sigma^{2}}\right]
\nonumber\\
& \mbox{} &
\times
J_{0}\left(\frac{\omega r^{\prime}r}{z-z^{\prime}}\right)
\exp\left[\frac{\I\omega(r^{\prime})^{2} +
\I\omega r^{2}}
{2c(z-z^{\prime})}\right] \ .
\label{eq:3}
\end{eqnarray}

\noindent It is convenient to rewrite this expression in a
dimensionless form.  After an appropriate normalization it is a function
of one dimensionless parameter only:

\begin{eqnarray}
& \mbox{} &
\hat{E} = f(\hat{z},\hat{r},N) =
\frac{\I}{N}\int\limits^{\hat{z}}_{0}\frac{\D\hat{z}^{\prime}}
{\hat{z}-\hat{z}^{\prime}}\int\limits^{\infty}_{0}
\D\hat{r}^{\prime}
\hat{r}^{\prime}\exp\left[-\frac{(\hat{r}^{\prime})^{2}}
{2N}\right]
\nonumber\\
& \mbox{} &
\times
J_{0}\left(\frac{
\hat{r}^{\prime}\hat{r}}{\hat{z}-\hat{z}^{\prime}}\right)
\exp\left[\frac{\I (\hat{r}^{\prime})^{2} +
\I \hat{r}^{2}}
{2(\hat{z}-\hat{z}^{\prime})}\right] \ .
\label{eq:4}
\end{eqnarray}

\noindent where $\hat{z} = z/L_{\mathrm{w}}$, $\hat{r} =
\sqrt{kr^{2}/L_{\mathrm{w}}}$, $k = \omega/c$, $L_{\mathrm{w}}$ is the
total undulator length, $N = k\sigma^{2}/L_{\mathrm{w}}$ is the
diffraction parameter (or, Fresnel number of the electron beam),
$\hat{E} = \tilde{E}/E_{0}$ is the normalized field amplitude, and

\begin{displaymath}
E_{0} =
\frac{\theta_{\mathrm{w}}\omega a_{\mathrm{in}}I_{0}}
{2c^{2}} \ .
\end{displaymath}

\noindent Integrating first with respect to $\hat{r}^{\prime}$, we have

\begin{equation}
\hat{E} = \I\int\limits^{\hat{z}}_{0}
\frac{\D\hat{z}^{\prime}}{\hat{z}-\hat{z}^{\prime} + \I N}
\exp\left[ - \frac{\I\hat{r}^{2}}{2(\hat{z} - \hat{z}^{\prime} + \I
N)}\right] \ .
\label{eq:nf}
\end{equation}

\noindent The integration over source coordinate $\hat{z}^{\prime}$ can
be performed without great difficulty in limiting case, namely, the
case of diffraction parameter very large compared with unity. In this
case the integral in (\ref{eq:nf}) is calculated analytically

\begin{displaymath}
\hat{E} = \frac{\hat{z}}{N}\exp\left( - \frac{\hat{r}^{2}}{2N}\right)
\qquad {\mathrm{as}}\quad N \gg 1 \ .
\end{displaymath}

\noindent It is convenient to express electric field inside the wide
electron beam in dimension form

\begin{equation}
\tilde{E}(z,\vec{r}_{\perp}) = \pi e\theta_{\mathrm{w}}za_{\mathrm{in}}
n_{0}(\vec{r}_{\perp}) = \frac{\theta_{\mathrm{w}}za_{\mathrm{in}}
I_{0}}{2c\sigma^{2}}\exp\left( - \frac{r^{2}}{2\sigma^{2}}\right)
\quad {\mathrm{as}}\quad N \gg 1 \ .
\label{eq:wba}
\end{equation}

\noindent Note that this result is completely general: that is, it
applies for any electron beam profile.  To calculate equation
(\ref{eq:a8}) we note that the behavior of Green's function for
$k\sigma^{2} \gg L_{\mathrm{w}}$ approaches the behavior of the delta
function. The source function $n_{0}(\vec{r}^{\prime}_{\perp})$ does
not vary very much across the region $\mid\vec{r}_{\perp} -
\vec{r}^{\prime}_{\perp}\mid^{2} \simeq L_{\mathrm{w}}/k$ in the case
of wide electron beam: therefore we can replace it by a constant. In
other words, we simply take $n_{0}(\vec{r}^{\prime}_{\perp})$ outside
the integral sign and call it $n_{0}(\vec{r}_{\perp})$.

\begin{figure}[tb]
\begin{center}
\epsfig{file=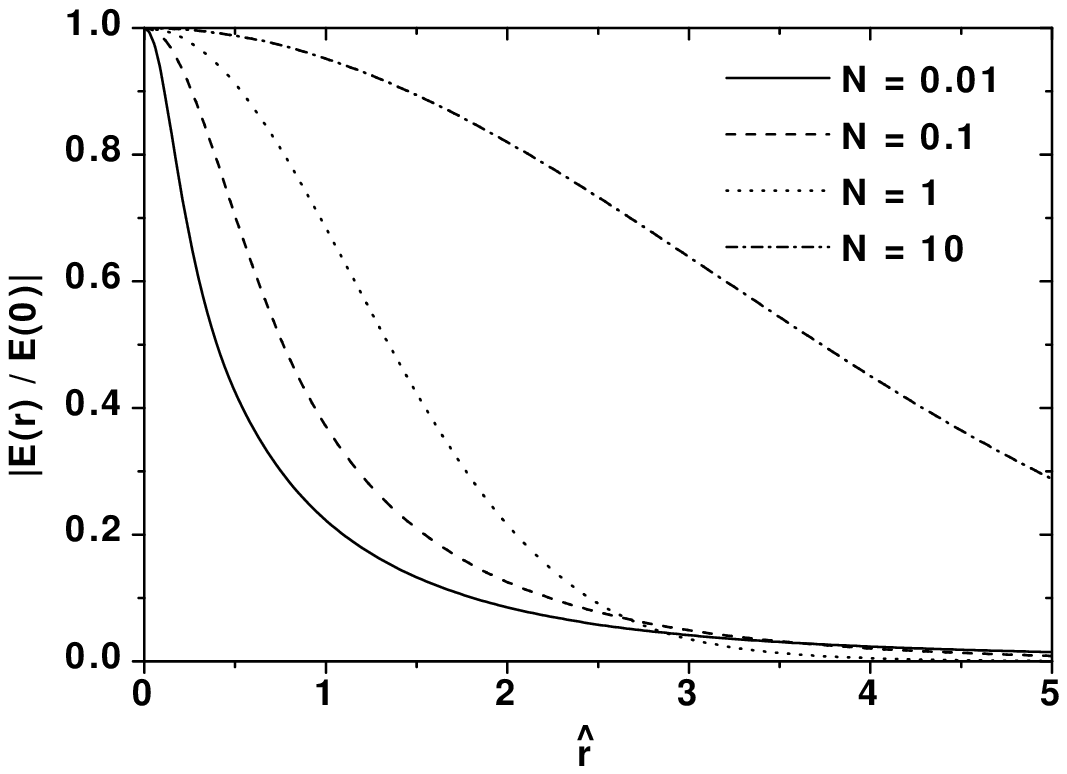,width=0.75\textwidth}

\vspace*{-10mm}

\epsfig{file=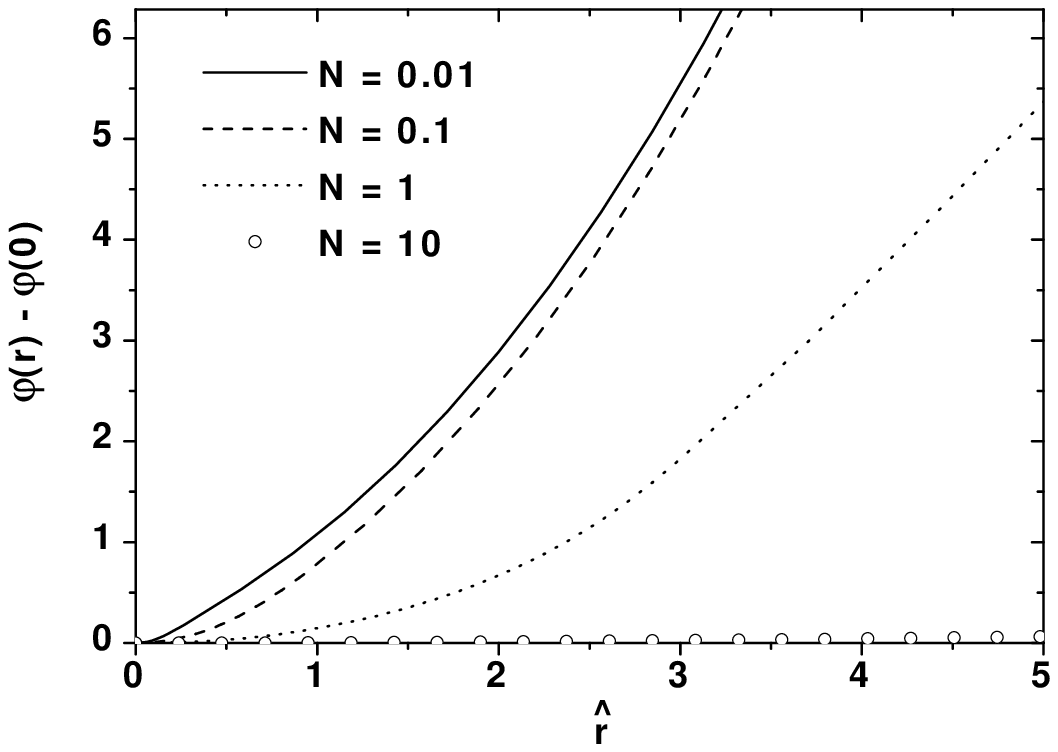,width=0.75\textwidth}
\end{center}
\caption{
Electron beam with Gaussian profile: transverse distribution
of the field amplitude (top) and phase (bottom)
at the radiator undulator exit for several
values of the diffraction parameter $N$ from 0.01 to 10.  Here detuning
parameter $C = 0$
}
\label{fig:nearze}
\end{figure}


The radiation field distribution at the exit of undulator is one of the
important characteristics of the radiator. For the case of Gaussian
electron beam profile transverse profile of the radiation field is
presented in Fig.~\ref{fig:nearze}.  Since the slow varying field
amplitude $\hat{E}(\hat{z}, \hat{r})$ is given by complex function of
the transverse coordinate, the wavefront of output radiation is not
plane.  It is interesting to trace the variation of the field phase
across the radiation beam. Bottom plot in Fig. \ref{fig:nearze} shows
the distribution of the phase of the radiation beam for several values
of the diffraction parameter $N$.

\begin{figure}[tb]
\begin{center}
\epsfig{file=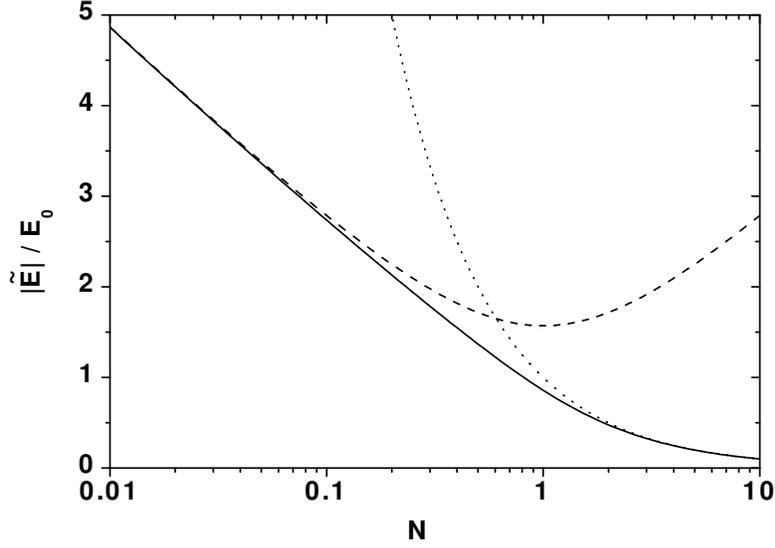,width=0.75\textwidth}
\end{center}
\caption{
Electron beam with Gaussian profile. The reduced radiation
field amplitude versus the diffraction parameter at $\hat{z} = 1,
\hat{r} = 0$.  Here detuning parameter $C = 0$. Dashed line shows
asymptote of thin electron beam. Dotted line shows asymptote of wide
electron beam
}
\label{fig:fzb0}
\end{figure}

It is interesting to plot the normalized amplitude of electric field as
a function of diffraction parameter in order to see how sensitive it is
to electron beam size. At this point we find it convenient to impose
the following restriction: we focus only on the radiation seen by
observer lying on the electron beam axis. We show such a plot in Fig.
\ref{fig:fzb0}.  When $\hat{r} = 0$ we can write (\ref{eq:nf}) in the
form

\begin{displaymath}
\hat{E} = f(\hat{z},0,N) =
\I\int\limits^{\hat{z}}_{0}
\frac{\D\hat{z}^{\prime}}{\hat{z}-\hat{z}^{\prime} + \I N}
= \arctan\left(\frac{\hat{z}}{N}\right) + \frac{\I}{2}\ln\left(1 +
\frac{\hat{z}^{2}}{N^{2}}\right)  \ .
\end{displaymath}

\noindent Let us study the asymptotic behavior of the field amplitude
at the large values of the diffraction parameter $N$. In this case
$\hat{z}/N \ll 1$ and we have asymptotically:

\begin{displaymath}
\hat{E} = f(\hat{z},0,N) \to \hat{z}/N \qquad {\mathrm{as}}
\quad N \to \infty \ .
\end{displaymath}

\noindent Now let us study the asymptote of a thin electron beam. In
this case $N \to 0$ and the function $f(\hat{z},0,N)$ can be estimated
simply as:

\begin{equation}
\hat{E} = f(\hat{z},0,N) \to \pi/2 + \I\ln(\hat{z}/N) \qquad
{\mathrm{as}} \quad N \to 0 \ .
\label{eq:log}
\end{equation}

\noindent Special attention is called to the fact that in the thin beam
case, at $N \to 0$, amplitude $\tilde{E}$ is a complex function. One
immediately recognizes the physical meaning of the complex $\tilde{E}$.
Note that electric field (response) is given by the fast wave of
transverse current ("force") times a certain factor. This factor can
either be written as $p+\I q$, or as certain magnitude $\rho$ times
$\exp(\I\delta)$. If it is written as a certain amplitude $\rho$ times
$\exp(\I\delta)$, let us see what it means. This tells us that electric
field is not oscillating in phase with the fast wave of transverse
current, which has (at $C = 0$) the phase $\psi = \omega z/c - \omega
t$, but is shifted by an extra amount $\delta$. Therefore $\delta(z)$
represent the phase shift of the response. For the experts in FEL
physics we should add that logarithmic terms in (\ref{eq:log}) and
logarithmic growth rate asymptote for conventional FEL amplifier at
small diffraction parameter (see \cite{book}) are ultimately connected.

From practical point of view it is necessary to know the field
distribution in the space after the undulator, at $z > L_{\mathrm{w}}$.
When the radiation field leaves the undulator, it is subjected to the
parabolic equation

\begin{displaymath}
\left(\vec{\nabla}^{2}_{\perp}
+ \frac{2\I\omega}{c}\frac{\partial}{\partial z}\right)\tilde{E}
= 0 \ .
\end{displaymath}

\noindent It follows from the latter equation that the field
amplitude in the space after the undulator and the field amplitude at
the undulator exit are connected by

\begin{displaymath}
\tilde{E}(z,\vec{r}_{\perp}) =
\frac{1}{4\pi(z-L_{\mathrm{w}})}\int\D\vec{r}^{\prime}_{\perp}
\tilde{E}(L_{\mathrm{w}},
\vec{r}^{\prime}_{\perp})\exp\left[\frac{\I\omega\mid\vec{r}_{\perp}
-\vec{r}^{\prime}_{\perp}\mid^{2}}{2c(z-L_{\mathrm{w}})}\right] \ .
\end{displaymath}

The subject of particular interest is the angular distribution of the
radiation intensity. The radiation field at the undulator exit may be
presented as a superposition of plane waves, all with the same wave
number $k = \omega/c$. The value of $k_{\perp}/k$ gives the sine of the
angle between the $z$ axis and the direction of propagation of the
plane wave (we consider the axisymmetric case). In the paraxial
approximation $k_{\perp}/k = \sin\theta \simeq \theta$. The angular
distribution of the radiation intensity, $I(\theta)$, can be expressed
as follows:

\begin{displaymath}
\frac{I(\theta)}{I(0)} =
\frac{\mid\Xi(\theta)\mid^{2}}{\mid\Xi(0)\mid^{2}} \ ,
\end{displaymath}

\noindent where $\Xi(\theta)$ is the spatial Fourier transform of the
complex amplitude of the radiation field, $\tilde{E}(z,r)$ at
the exit of the undulator. In the axisymmetric case the spatial Fourier
transform of the radiation field is given by

\begin{displaymath}
\Xi(\hat{\theta}, N) =
\int\limits^{\infty}_{0}\hat{E}(1,\hat{r}, N)
J_{0}(\hat{\theta}\hat{r})\hat{r}\D\hat{r} \ ,
\end{displaymath}

\noindent where $\hat{\theta} = \sqrt{kL_{\mathrm{w}}}\theta$ and
$J_{0}$ is the Bessel function of the first kind. Using (\ref{eq:nf}),
we find the expression for the angular distribution of the radiation
intensity,

\begin{displaymath}
\frac{I(\theta)}{I(0)} =
\frac{\mid\Xi(\theta)\mid^{2}}{\mid\Xi(0)\mid^{2}}
= \left[\frac{\sin\hat{\theta}^{2}/4}
{\hat{\theta}^{2}/4}\right]^{2}
\exp\left( - N\hat{\theta}^{2}\right) \ .
\end{displaymath}

\begin{figure}[tb]
\begin{center}
\epsfig{file=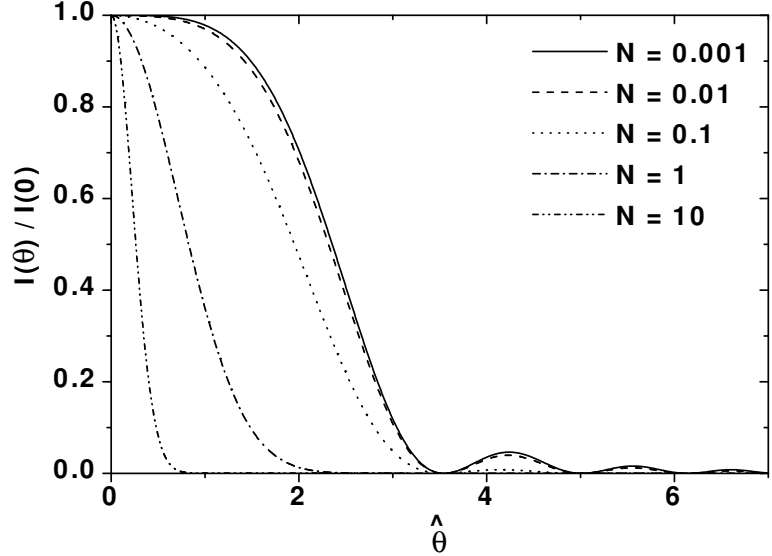,width=0.75\textwidth}
\end{center}
\caption{
Electron beam with Gaussian profile. Angular distribution of
the radiation intensity for several values of the diffraction parameter
$N$ from 0.001 to 10. Here detuning parameter $C = 0$
}
\label{fig:farz}
\end{figure}

\noindent At large value of diffraction parameter $N$ the far zone
approximation may be used when $z_{0}/(k\sigma^{2}) \gg 1$, where
$z_{0}$ is the distance between the observation point and the undulator
exit. When $N < 1$ the above condition changes to $z_{0} \gg
L_{\mathrm{w}}$. Figure \ref{fig:farz} presents the angular distribution
of the radiation intensity for the coherent undulator radiation in the
case of Gaussian profile of electron beam.  One can see that the
radiation power is mainly concentrated in the small angle near the $z$
axis. At large value of the diffraction parameter the distribution is
approximately equal to

\begin{displaymath}
\frac{\mid\Xi(\hat{\theta})\mid^{2}}{\mid\Xi(0)\mid^{2}}
\to \exp(-N\hat{\theta}^{2})
\qquad {\mathrm{as}} \quad N \to \infty \ .
\end{displaymath}

\noindent In the case of a thin electron beam  we have asymptotically:

\begin{displaymath}
\frac{\mid\Xi(\hat{\theta})\mid^{2}}{\mid\Xi(0)\mid^{2}}
\to \left[\frac{\sin\hat{\theta}^{2}/4}
{\hat{\theta}^{2}/4}\right]^{2}
\qquad {\mathrm{as}} \quad N \to 0 \ .
\end{displaymath}

Let us investigate qualitatively the process of coherent undulator
radiation. To get an intuitive picture on what happens with the
radiation beam let us first choose a thin beam asymptotic. This is an
example in which diffraction effects play an important role.  Simple
physical consideration can lead directly to a crude approximation for
the radiation beam cross-section. There is a complete analogy between
the radiation effects of the bunched electron beam in the undulator and
the radiation effects of a sequence of periodically spaced oscillators.
The radiation of these oscillators always interferes coherently at zero
angle with respect to the undulator axis.  When all the oscillators are
in phase there is a strong intensity in the direction $\theta = 0$. An
interesting question is, where is the minimum? If we have a triangle
with a small altitude $r \simeq z\theta$ and a long base $z$, then the
diagonal $s$ is longer than the base. The difference is $\Delta = s-z
\simeq r^{2}/2z \simeq z\theta^{2}/2$.  When $\Delta$ is equal to one
wavelength, we get a minimum because the contributions of various
oscillators are then uniformly distributed in phase from $0$ to $2\pi$.
In the limit of small size of the electron beam interference will be
constructive within an angle of about $\theta_{\mathrm{c}} \simeq
(\sqrt{kz})^{-1}$. In the limit of large electron beam size, the field
is concentrated mainly inside the electron beam.  The radiation field
across the electron beam may be present as a superposition of plane
waves. We can expect that the typical width of the angular spectrum
should be of the order $\theta_{\mathrm{c}} \simeq (k\sigma)^{-1}$,
simply a consequence of the reciprocal width relations of the Fourier
transform pair $\Delta k_{\perp}\sigma \simeq 1$.

The boundary between these two asymptotes is about $ k\sigma \simeq
\sqrt{kL_{\mathrm{w}}}$ or (another way to write it) $\sigma^{2} \simeq
\sigma^{2}_{\mathrm{dif}} = L_{\mathrm{w}}/k$. A rough estimate for the
diffraction effects to be small is $\sigma \gg
L_{\mathrm{w}}/(k\sigma)$, which simply means that the diffraction
expansion of the radiation at undulator length must be much less than
the size of the beam. Another way to write this condition is
$k\sigma^{2}/L_{\mathrm{w}} = N \gg 1$. As we mentioned above, the
diffraction parameter $N$ can be referred to as the electron beam
Fresnel number.

Let us consider the electromagnetic power. The well-known Poynting
vector represents the electromagnetic power flow. In the paraxial
approximation the diffraction angles are small, the vectors of electric
and magnetic field are equal in absolute value and are perpendicular to
each other. Thus, the expression for the radiation power, $W$, can be
written in the form:

\begin{equation}
W =
\frac{c}{4\pi}\int\overline{\mid\vec{E}_{\perp}\mid^{2}}
\D\vec{r}_{\perp}  =
\frac{c}{4\pi}\int\mid\tilde{E}(z,\vec{r}_{\perp})\mid^{2}
\D\vec{r}_{\perp}  \ ,
\label{eq:po}
\end{equation}

\noindent where $\overline{(\cdots)}$ denotes averaging over a cycle of
oscillation of the carrier wave. If we consider a system with fields
and bunched electron beam in an undulator, the energy stored in any
volume fluctuates sinusoidally with time. But on the average there is
no increase or decrease in the energy stored in any portion of the
volume.

Now we shall calculate the output power. To determine the
$W$ we substitute (\ref{eq:3}) into (\ref{eq:po}). It is convenient to
write the expression for $W$ in a dimensionless form.  After
an appropriate normalization it is a function of one dimensionless
parameter only:

\begin{equation}
\hat{W} = F(N) \ ,
\label{eq:wh}
\end{equation}

\noindent where $N = k\sigma^{2}/L_{\mathrm{w}}$  is the diffraction
parameter, $\hat{W} = W/W_{0}$ is the normalized power, and $W_{0}$ is

\begin{displaymath}
W_{0} =
\frac{\pi\theta^{2}_{\mathrm{w}}\omega I_{0}^{2}a^{2}_{\mathrm{in}}
L_{\mathrm{w}}}
{8c^{2}} \ .
\end{displaymath}

\noindent Substituting the expression for $\tilde{E}$ from
(\ref{eq:nf}) into (\ref{eq:po}), we obtain:

\begin{equation}
F(N) = \frac{2}{\pi}\left[\arctan\left(\frac{1}{2N}\right)
+ N\ln\left(\frac{4N^{2}}{4N^{2}+1}\right)\right] \ .
\label{eq:f}
\end{equation}

\begin{figure}[tb]
\begin{center}
\epsfig{file=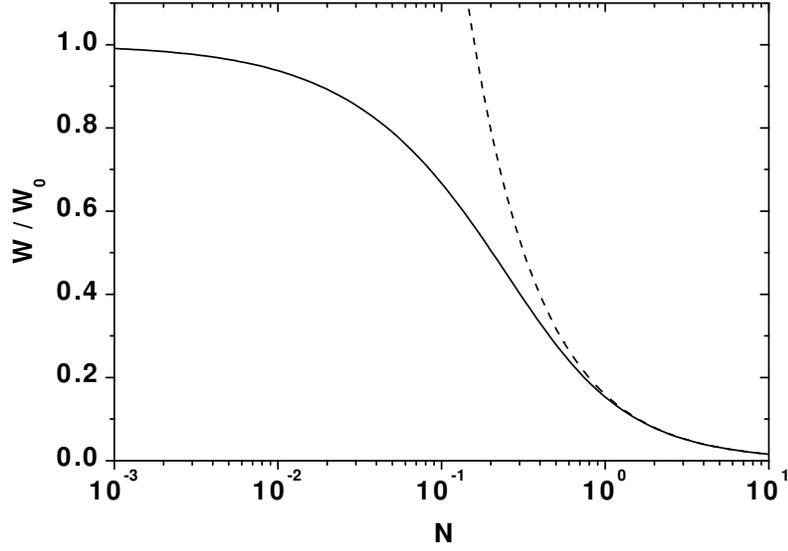,width=0.75\textwidth}
\end{center}
\caption{
Electron beam with Gaussian profile. The reduced output power versus
the diffraction parameter. Solid curve is calculated with analytical
formula (\ref{eq:f}). Dashed curve shows an asymptote for a wide
electron beam. Here the detuning  parameter is $C = 0$
}
\label{fig:pcohu}
\end{figure}

\noindent In Fig. \ref{fig:pcohu} we present a plot of this universal
function.  The physical implication of this result are best understood
by considering some limiting cases. We have asymptotically:

\begin{displaymath}
F(N) \to 1/(2\pi N) \quad {\mathrm{as}} \qquad N \to \infty \ ,
\end{displaymath}

\begin{displaymath}
F(N) \to 1 \quad {\mathrm{as}} \qquad N \to 0 \ ,
\end{displaymath}

\noindent Let  us notice a remarkable feature of that plot. In the
limit of a thin electron beam, $N \to 0$, the radiation power tends to
a constant value $W \to W_{0}$ and the dependence of output radiation
on the transverse size of the electron beam is rather weak.

For practical purposes it is convenient to express $W_{0}$ in
an explicit form:

\begin{equation}
W_{0} =
W_{\mathrm{b}}\left[\frac{\pi^{2}a^{2}_{\mathrm{in}}}{2}\right]\left[
\frac{I}{\gamma I_{\mathrm{A}}}\right]\left[\frac{K^{2}}{1 +
K^{2}}\right]N_{\mathrm{w}} \ ,
\label{eq:w0}
\end{equation}

\noindent where $W_{\mathrm{b}} = m_{\mathrm{e}}c^{2}\gamma I_{0}/e$ is
the total power of electron beam, $I_{\mathrm{A}} =
m_{\mathrm{e}}c^{3}/e \simeq 17$ kA is the Alfven current.
Let us make a calculation of $W$ for some cases. Suppose
$a_{\mathrm{in}} = 0.3$, $I_{0} = 3$ kA, $\gamma = 10^{3}$, $K = 5.4$,
$N_{\mathrm{w}} = 5$; then by equation (\ref{eq:w0}) it follows that
$W_{0} \simeq 500$ MW.  If the laser wavelength $\lambda = 1 \mu$m,
the normalized transverse emittance $\epsilon_{\mathrm{n}} = 2\pi\mu$m,
focusing beta function is equal to 1 m, the diffraction parameter is
about 0.04.  Remembering the previous result (see Fig. \ref{fig:pcohu})
we come to the conclusion that we can treat this situation as a
coherent undulator radiation with thin electron beam.

It is relevant to make some remarks on the region of applicability of
the results of this section. One of the basic assumptions of the theory
is that  the radius of the electron rotation in the undulator,
$r_{\mathrm{w}} = \theta_{\mathrm{w}}/k_{\mathrm{w}}$, is much less
than the transverse size of the electron beam. Taking into account that
$\theta_{\mathrm{w}} = K/\gamma$, we can write

\begin{displaymath}
\frac{\sigma^{2}}{r^{2}_{\mathrm{w}}}
=
\frac{1+K^{2}}{K^{2}}\left(\gamma^{2}_{z}
\sigma^{2}k^{2}_{\mathrm{w}}\right)  = \frac{1+K^{2}}{K^{2}}\left(\pi
N_{\mathrm{w}}N\right) \gg 1 \ .
\end{displaymath}

\noindent Thus, the requirement for the parameter
$\sigma^{2}/r^{2}_{\mathrm{w}}$ to be large can be written as $\pi
N_{\mathrm{w}}N \gg 1$. When the diffraction parameter $N$ is much
larger than $(\pi N_{\mathrm{w}})^{-1}$, the radius of the electron
rotation in the undulator is always much less than the transverse size
of the electron beam. In our example we have $N \simeq (\pi
N_{\mathrm_{w}})^{-1}$. We should say that this particular case is at
the boundary of the region of the applicability of our theory.

The formula for the output power which we derived (\ref{eq:wh}) refer
to the case of the helical undulator. It can be simply generalized for
the case of a planar undulator. The magnetic field on the axis of the
planar undulator is given by

\begin{displaymath}
\vec{H} = \vec{e}_{x}H_{\mathrm{w}}\cos(k_{\mathrm{w}}z) \ .
\end{displaymath}

\noindent The explicit expression for the electron velocity in the field
of the planar undulator has the form:

\begin{displaymath}
\vec{v}_{\perp} = -
\vec{e}_{y}c\theta _{\mathrm{w}}\sin(k_{\mathrm{w}}z) \ ,
\end{displaymath}

\noindent where
$\theta_{\mathrm{w}} = K/\gamma =
\lambda_{\mathrm{w}}eH_{\mathrm{w}}/(2\pi m_{\mathrm{e}}c^{2}\gamma)$.
The constrained motion of the electron in the planar undulator differs
from that in the helical one. An important feature of this motion is
that longitudinal velocity $v_{z}$ of the electron oscillates along the
undulator axis which creates definite problems for the description of
the radiation process. It is not hard to go through the derivation of
radiation power again. If we do that, and calculate the power in the same
way, we get that all the expressions for the planar undulator are
identical to those for the helical undulator. The only difference is
the appearance of different numerical factors taking their origin from
the averaging procedure. One can obtain that expression for power
written down in the reduced form is identical for both undulator
configurations. As a result, the universal plot $\hat{W} = F(N)$ in
Fig. \ref{fig:pcohu} is applicable to the case of planar undulator, too.
The only difference is that the following definition of normalization
factor for the radiator with planar undulator should be used:

\begin{displaymath}
W_{0} =
\frac{\pi\theta^{2}_{\mathrm{w}}A^{2}_{\mathrm{JJ}}\omega_{0}
I_{0}^{2}a^{2}_{\mathrm{in}} L_{\mathrm{w}}} \ ,
{16c^{2}}
\end{displaymath}

\noindent where

\begin{displaymath}
A_{\mathrm{JJ}} = [J_{0}(Q) - J_{1}(Q)] \ ,
\end{displaymath}

\noindent $J_{n}(Q)$ is a Bessel function of $n$th order, and

\begin{displaymath}
Q = \theta^{2}_{\mathrm{w}}\omega_{0}/(8k_{\mathrm{w}}\gamma^{2})
= K^{2}/(4 + 2K^{2}) \ .
\end{displaymath}

\noindent When we simplified the expression for $Q$, we used the
resonance condition for the planar undulator $\omega_{0} =
2\gamma^{2}k_{\mathrm{w}}/[c(1+K^{2}/2)]$. For practical purposes it is
convenient to rewrite the expression for $W_{0}$ in the form

\begin{displaymath}
W_{0} =
W_{\mathrm{b}}\left[\frac{\pi^{2}a^{2}_{\mathrm{in}}}{2}\right]\left[
\frac{I}{\gamma I_{\mathrm{A}}}\right]\left[\frac{K^{2}}{2 +
K^{2}}\right]A^{2}_{\mathrm{JJ}}N_{\mathrm{w}} \ .
\end{displaymath}

Let us present a specific numerical example for the case of a radiator
with a planar undulator. With the numerical values $\lambda_{\mathrm{w}}
= 6.5$ cm, $K = 7.6$, $\gamma = 10^{3}$, the resonance value of
wavelength is $\lambda = 1 \mu$m. If the number of the undulator
periods $N_{\mathrm{w}} = 5$, the amplitude of density modulation
$a_{\mathrm{in}} = 0.3$, the beam peak current $I_{0} = 3$ kA, the
radiation power is about $W = W_{0}F(N) \simeq 250 F(N)$ MW.

\begin{figure}[tb]
\begin{center}
\epsfig{file=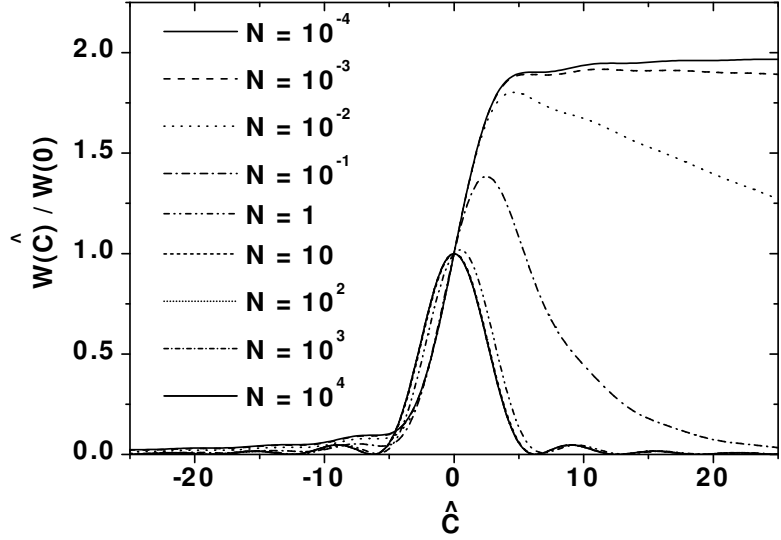,width=0.75\textwidth}
\end{center}
\caption{
Electron beam with Gaussian profile. The reduced output power versus
the reduced detuning $\hat{C}$ for various values of Fresnel number
}
\label{fig:pcohdet}
\end{figure}

\begin{figure}[tb]

\epsfig{file=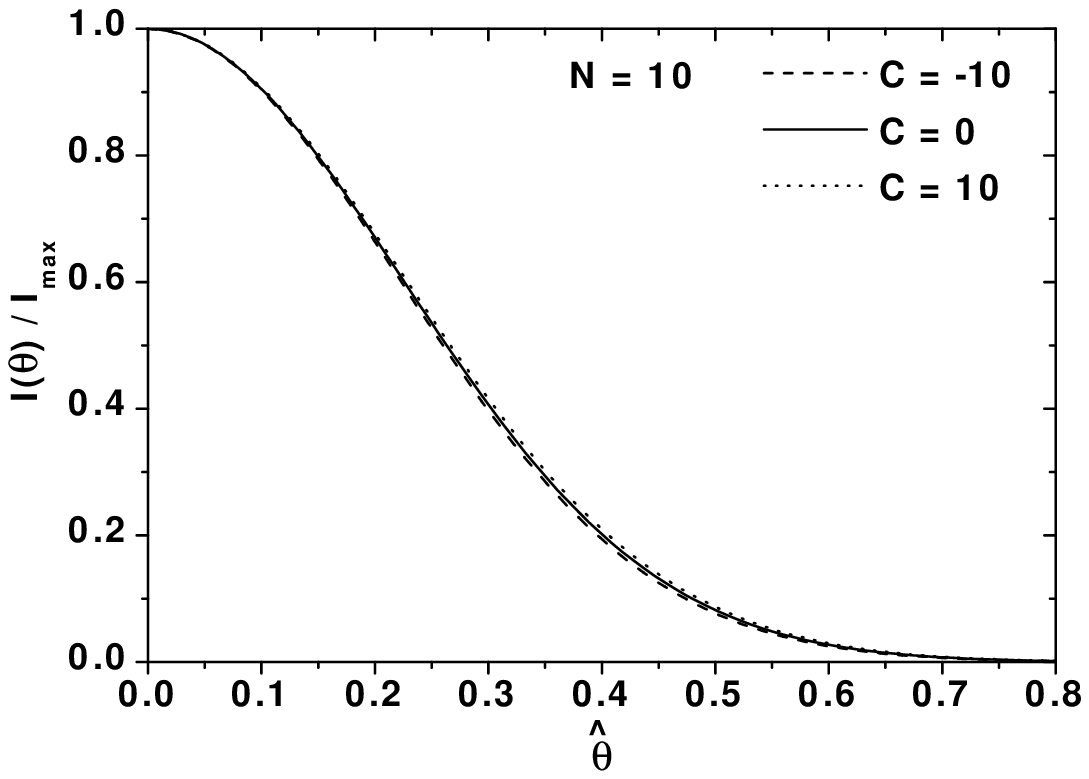,width=0.5\textwidth}

\vspace*{-61mm}

\hspace*{0.5\textwidth}
\epsfig{file=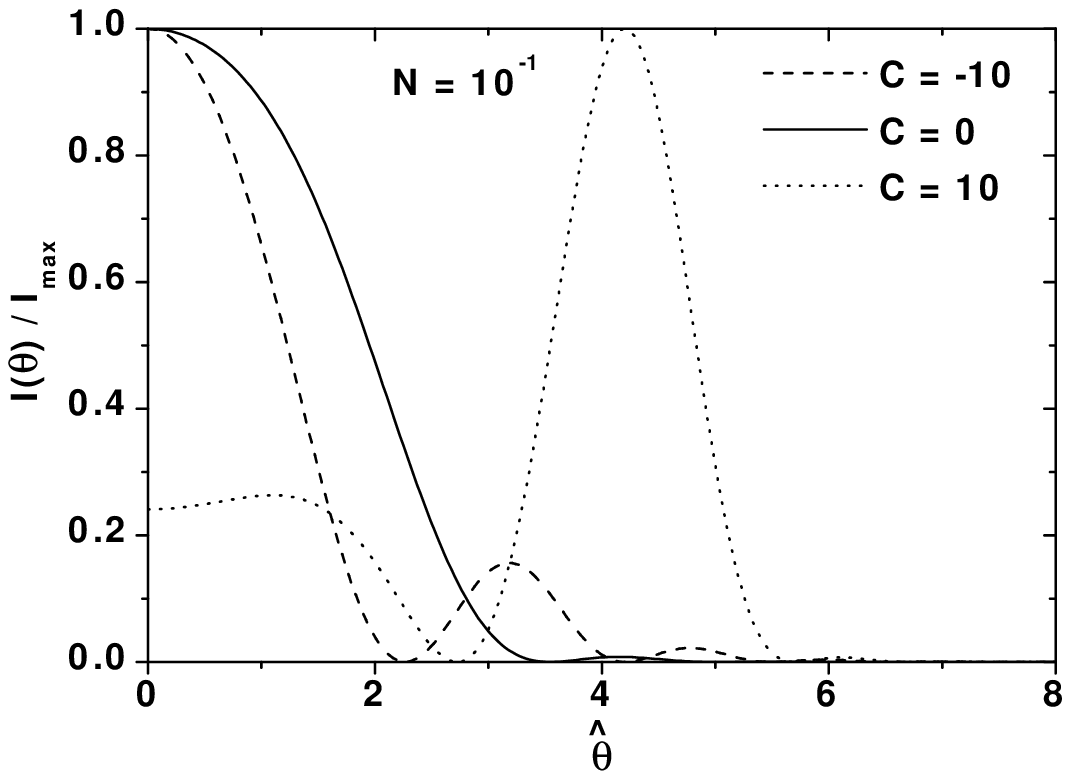,width=0.5\textwidth}

\epsfig{file=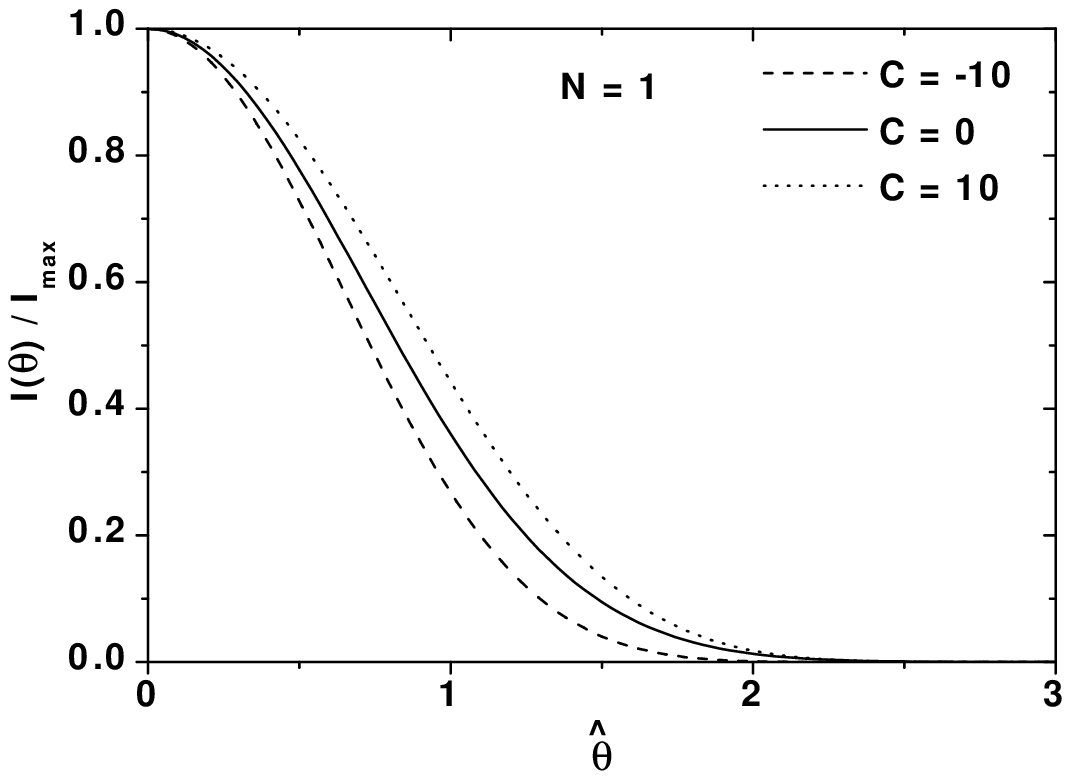,width=0.5\textwidth}

\vspace*{-61mm}

\hspace*{0.5\textwidth}
\epsfig{file=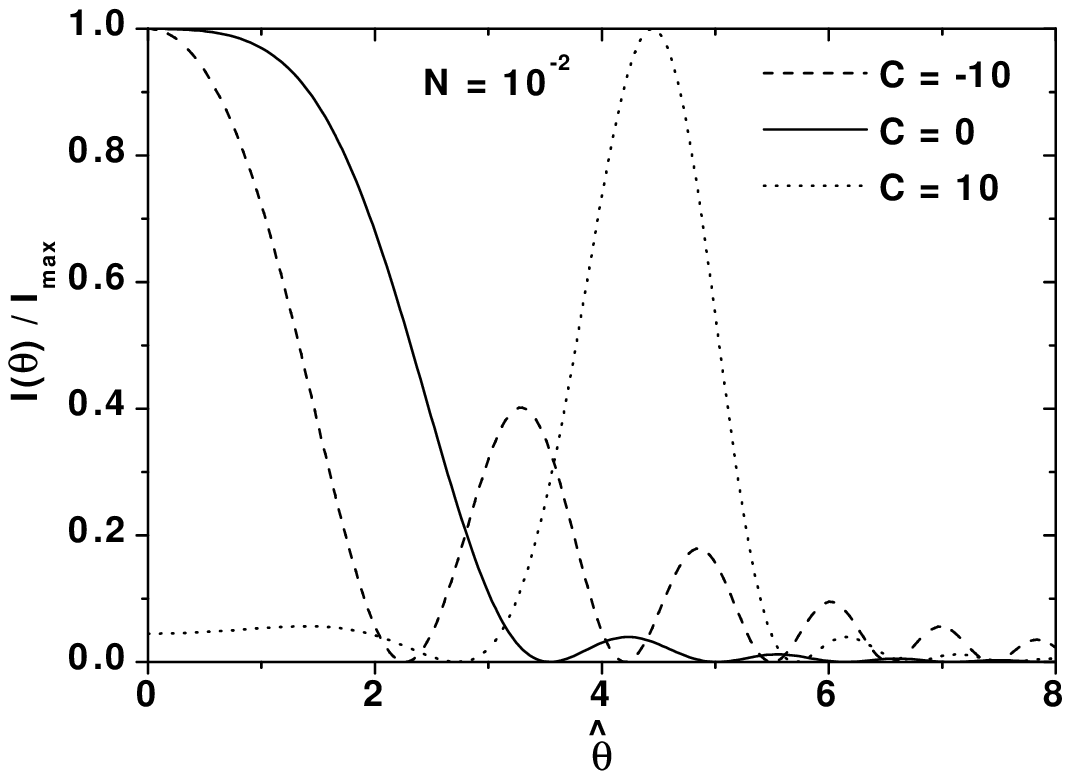,width=0.5\textwidth}




\caption{
Electron beam with Gaussian profile. Angular distribution of the
radiation intensity for several values of the diffraction parameter.
Curves correspond to the detuning parameter $\hat{C} = -10, 0$, and 10
}
\label{fig:fzdet1}
\end{figure}

All of the foregoing discussion of coherent undulator radiation has
been concerned solely for the radiation at resonance -- that is $\omega
= \omega_{0} = 4\pi\gamma^{2}_{z}c/\lambda_{\mathrm{w}}$. Now, we would
like to find out how the output radiation varies in the circumstance
that seed signal frequency $\omega$ is nearly, but not exactly, equal
to $\omega_{0}$. According to the radiation equation (\ref{eq:a7}), the
radiation process is determined by the detuning $C = k_{\mathrm{w}} +
\omega/c - \omega/v_{z}$ which is the function of the seed laser
frequency, energy of the electron beam and the undulator parameter. It
is not hard to go through the derivation of output radiation power
again. If we take $C \ne 0$, the solution of the wave equation
(\ref{eq:a7}) has the form

\begin{eqnarray}
& \mbox{} &
\tilde{E}(z,\vec{r}_{\perp}) = \frac{\I
e\theta_{\mathrm{w}}\omega a_{\mathrm{in}}}
{2c}\int\limits^{z}_{0}\frac{\D
z^{\prime}}{z-z^{\prime}}\exp(-\I Cz^{\prime})
\nonumber\\
& \mbox{} &
\times
\int\D\vec{r}_{\perp}n_{0}
(\vec{r}_{\perp}^{\prime})
\exp\left[\frac{\I\omega\mid\vec{r}_{\perp}
-\vec{r}^{\prime}_{\perp}\mid^{2}}{2c(z-z^{\prime})}\right] \ .
\label{eq:6}
\end{eqnarray}

\noindent When the electron beam profile is Gaussian, we can write
(\ref{eq:6}) in a dimensionless form:

\begin{eqnarray}
& \mbox{} &
\hat{E} = f(\hat{z},\hat{r},N,\hat{C}) =
\frac{\I}{N}\int\limits^{\hat{z}}_{0}\frac{\D\hat{z}^{\prime}}
{\hat{z}-\hat{z}^{\prime}}\int\limits^{\infty}_{0}
\D\hat{r}^{\prime}
\hat{r}^{\prime}\exp\left[-\frac{(\hat{r}^{\prime})^{2}}
{2N}\right]\exp(-\I\hat{C}\hat{z}^{\prime})
\nonumber\\
& \mbox{} &
\times
J_{0}\left(\frac{
\hat{r}^{\prime}\hat{r}}{\hat{z}-\hat{z}^{\prime}}\right)
\exp\left[\frac{\I (\hat{r}^{\prime})^{2} +
\I \hat{r}^{2}}
{2(\hat{z}-\hat{z}^{\prime})}\right]  = \I\int\limits^{\hat{z}}_{0}
\frac{\D\hat{z}^{\prime}}{\hat{z}-\hat{z}^{\prime} + \I N}
\nonumber\\
& \mbox{} &
\times
\exp\left[ - \frac{\I\hat{r}^{2}}{2(\hat{z} - \hat{z}^{\prime} + \I
N)}\right]\exp\left(-\I\hat{C}\hat{z}^{\prime}\right) \ .
\label{eq:7}
\end{eqnarray}

\noindent We use the notations similar to those introduced above.
Also, an additional parameter of the problem, the dimensionless
detuning parameter $\hat{C} = CL_{\mathrm{w}}$, appears in the theory,
since we take into account resonance behavior. Let us express
$\hat{C}$ in terms of physical parameters. The detuning
parameter $\hat{C}$ is connected by the simple relation with the
frequency deviation:  $\omega - \omega_{0} = \Delta\omega = -
2\gamma^{2}_{z}C$. Thus, we obtain $\hat{C} = - 2\pi N_{\mathrm{w}}
\Delta\omega/\omega_{0}$, where $N_{\mathrm{w}}$ is the number of
radiator undulator periods.

Let us now study the influence of the detuning on the radiation
process. In Fig. \ref{fig:pcohdet} the output power is shown as a
function of reduced detuning for different values of diffraction
parameter.  One can see that the radiation process displays resonance
behavior and the output power depends strongly on the value of
the detuning parameter $\hat{C}$. It is seen from the plot that at
large value of Fresnel number the resonance curve is simply that of the
interference factor,

\begin{displaymath}
f(\hat{C}) = \frac{\sin^{2}(\hat{C}/2)}{(\hat{C}/2)^{2}} \ .
\end{displaymath}

\noindent  One can see that this formula works well at $N \simeq 10$.
Then, at $N \simeq 1$, the resonance curve is visibly modified due to
diffraction effects. One can see from this plot that the resonance
curve is not completely symmetrical function of the detuning parameter
$\hat{C}$, the asymmetry being greater for smaller Fresnel number. The
reason for this is that at small values of diffraction parameter the
angular distribution of the radiation corresponds to that emitted by a
simple one electron. To explain this phenomena, we should analyze the
angular distribution of the radiation intensity. Even without
performing calculations, we can expect angular-frequency dependence for
the output radiation in the case of a thin electron beam asymptote. As
we can see from Fig. \ref{fig:fzdet1}, numerical calculations in the
far zone confirm this simple physical consideration.

\section{Discussion}

Successful operation of the ultrashort-pulse-measurement device (FROG)
requires the fulfillment of several requirements. The requirement for
the spatio-temporal pulse distortions to be small is of importance for
the performance of the FROG measurement apparatus. One of the problems
is the measurement of the pulses that have significant spatial
structure, for example, a pulse whose transverse size varies along the
pulse. For the FROG is of great interest to minimize ultrafast
variation of the optical replica transverse size which is due to
emittance variation along the electron bunch. Obviously, this
requirement is easier to achieve for a thin electron beam asymptote. On
the other hand, such ultrafast variation of the pulse transverse size
is an essential characteristic of the optical replica in a wide
electron beam case, since it reflects behavior of a slice emittance.
The obvious solution of this problem is as follows. A technique best
suited for the  pulses with spatial structure consists of expanding the
radiation beam and filtering out the central part to almost constant
transverse size. A spatial filter can be realized by using optical
arrangement where a telescope is placed between replica synthesizer
exit and FROG device. The pulse to be measured is propagated through
the hole which spatially filters the pulse. The important point is that
electric field of such filtered pulse contains completely the same
information about transverse size of the electron bunch as initial
pulse. If the electron beam is wide, $N \gg 1$, then electric field of
the filtered replica is inversely proportional to the square of
electron beam $E(t) \simeq {\mathrm{const.}}\times I(t)/\sigma^{2}(t)$.

One of the big unsolved problems of the electron bunch diagnostics is
measurement of bunches that have significant distortions in transverse
phase space, for example, a bunch whose transverse phase space ellipse
varies from point to point in the beam. We have considered in section 2
a simplified model of the electron beam and used the following
assumptions: i) the electron beam transverse profile is assumed to be
axisymmetric; ii) Twiss parameters are equal in all slices (although
emittances are different). Such a beam can, in principle, be realized
in an "idealized" RF photoinjector with a perfectly working emittance
compensation technique \cite{sem1} that allows one to align slices in
transverse phase space. For real beam, the variation in the space
charge forces can be significant and cannot be properly compensated
with solenoidal emittance compensation that was observed in different
measurements \cite{sem2,sem3,sem4}.  In addition, CSR-related effects
in bunch compressors can lead to further deviations from the simple
model.  It is clear that a knowledge of the variation of phase space
ellipse along the bunches at the output of the bunch formation system
could provide significant information about the physical mechanisms
responsible for generation of ultrashort bunches. Here we would like to
discuss a further extension of the proposed diagnostic  method that can
allow one to determine Twiss parameters in axial slices that are only a
$\mu$m-long fraction of the full bunch length.

A very simple approach, involving simultaneous "quadrupole-scan"  and
"hole-scan" techniques, yields the solution. The main idea can easily
be understood taking into account the fact that in the limit of a wide
electron beam we measure the beam size, and therefore, we can, in
principle, use a standard "quadrupole-scan" technique. Moreover, the
method of spatial filtering described above allows one to determine a
transverse distribution of the beam density in each slice. Indeed, in
the limit of a wide beam the electric field at the undulator exit
(\ref{eq:wba}) is directly proportional to the beam density
distribution (this holds also for non-axisymmetric beam when both
sizes, in $x-$ and $y-$directions, are much larger then a diffraction
size).  FROG technique and spatial filter allow the two-dimensional
slice density distribution measurements to be made using a "hole-scan"
technique. Therefore, one can do a quadrupole scan (using a system of
quadrupoles) by changing phase advance but keeping all slices in a wide
beam limit. For each settings of the quadrupoles one does
two-dimensional scan with a filter, checking that slice sizes are in
the limit of wide beam. Then emittances and Twiss parameters in each
slice are reconstructed by the standard method. If there are offsets
and angles of slice centroids (due to CSR effects, for example), they
can also be reconstructed from this measurement. Note also that after
two-dimensional scan of the hole one gets three-dimensional map of the
electric field amplitude that is a copy of three-dimensional density
distribution in the electron bunch. Since total charge is known, the
unknown absolute value of the current density can be determined, too.
This approach appropriately can be referred to as bunch phase space
tomography.

The main emphasis of previous considerations was concentrated on the
measurements of ultrashort (sub-100~fs long) electron bunches. The
method proposed can be also applied for measurements of long (a few ps)
bunches, too. Measurements of such bunches is practically important
problem. The X-ray FEL bunch compressors consist of a series of
magnetic chicanes. To setup the compression, the bunch structure needs
to be measured before and after each compression stage. In addition,
once the bunch compressors are set up, a bunch length feedback system
will be required for stabilization of the compression. The electron
pulse durations of interest are approximately 10, 2 and 0.2 ps,
respectively. The big advantage of the proposed diagnostic technique is
the absence of apparent limitations which would prevent determination
of the structure of electron bunches even without bunch compression in
the injector linac. According to our discussion above, the pulse energy
of optical replica is proportional to the value of the peak current
which is relatively weak dependence. The energy of radiated pulse is
about $10 \mu$J for the case when $I = 3$ kA. The energy per pulse is
decreased by only a factor 30, down to sub-$\mu $J level for $I =
100$~A. FROG technique still works well for such parameters of optical
pulse. It has been used to measure pulses from a few fs to many ps in
length. It has measured pulses from pJ to mJ in energy. FROG has proven
to be a general technique that works \cite{treb}.

Operation of the proposed scheme was illustrated for the parameters of
the European XFEL. Although the present work is illustrated for the
electron beam energy of 0.5 GeV, its applicability is not restricted to
this range. For example, LCLS bunch compressor system \cite{tdr2} is a
suitable candidate for application of diagnostic techniques described
here.

\section{Conclusion}

Sub-100 fs ultrarelativistic electron bunches, which only a few years
ago seemed like wishful thinking, are now effectively generated in the
accelerator test facilities and have given a rise to ultrafast X-ray
applications. The femtosecond time scale is beyond the range of
standard electronic display instrumentation, and the development of the
methods for the measurement of the longitudinal beam current
distribution in such short bunches is undoubtedly a challenging
problem. In this paper we proposed a new method for ultrashort electron
pulse-shape measurements. Making the use of the ultrashort laser
pulse-shape measurement device (FROG) together with carefully designed
undulator-based optical replica synthesizer allow the electron bunch
length measurement with resolution of about a few femtosecond. We
demonstrate that proposed measuring device can be used to determine the
electron current profile for a single ultrashort electron bunch, which
makes it an ideal online tool for optimization of complex bunch
compression systems. In general case the electron bunches may have
significant emittance and energy spread variation along the bunch. Proposed
device is capable to measure both of these electron bunch distortions
quantitatively, too. An important feature of the method is that all
steps of the optical replica synthesis are controlled by means of the
choice of the undulator parameters, dispersion section strength and
value of beta function. Data sets of beta function and dispersion
section strength scans actually contain all the required information
for retrieval of the slice properties of the bunch. Thus, proposed
technique combines full-information electron bunch measurement with
much-needed experimental simplicity. The only requirement for the
proposed technique is the capability of the electron bunch to generate
$1 \mu$m radiation, which implies a minimum on the electron energy of
about one hundred MeV. However, this limit nicely fits in the design
parameters of bunch compression systems for XFELs. Another key element
-- laser pulse-shape measurement device (FROG) -- is now a standard and
well-developed tool. All key elements of measuring device have already
been established. Operating range of proposed diagnostic technique
nicely includes that of most ultrashort X-ray FEL injector linacs, so
it should be ideal for most everyday diagnostics as well as many more
exotic ones.

\section*{Acknowledgments}
We thank G. Geloni for many useful discussions. We thank R.~Brinkmann,
J.R.~Schneider, A.~Schwarz, and D.~Trines for interest in this work.


\clearpage

\begin{thebibliography}{99}


\bibitem{ay1}
V.~Ayvazyan et al., Phys. Rev. Lett. 88(2002)104802.

\bibitem{ay2}
V.~Ayvazyan et al., Eur. Phys. J. D20(2002)149.

\bibitem{tdr1}
TESLA Technical Design Report, DESY2001-011, edited by Richard et al.,
and http://tesla.desy.de.

\bibitem{tdr-xfel}
P.~Audebert et al., ``TESLA XFEL: First stage of the X-ray laser
laboratory -- Technical design report (R.~Brinkmann et al., Eds.)'',
Preprint DESY 2002-167.

\bibitem{tdr2}
The LCLS Design Study Group, LCLS Design Study Report, SLAC reports
SLAC-R-593 (2002) and http://www-ssrl.slac.stanford.edu/lcls/CDR.


\bibitem{fp1}
M. Borland et al.,
Nucl. Instrum. and Methods A483(202)268.

\bibitem{fp2}
M. Dohlus et al., "Start-to-End Simulations of SASE FEL at the TESLA
Test Facility, Phase 1" Preprint DESY 03-197, DESY, Humburg, 2003;
Nucl. Instrum.  and Methods (in press).


\bibitem{emma}
P. Emma et al., Phys. Rev. Lett. 92(2004)074801.

\bibitem{treb}
R. Trebino, Frequency-Resolved Optical Gating: Measurement of
Ultrashort Laser Pulses, Kluwer Academic Publishers, Boston, 2002.


\bibitem{fast}
E.L. Saldin, E.A. Schneidmiller and M.V. Yurkov,
Nucl. Instrum. and Methods A429(1999)233.

\bibitem{ds1}
E.L. Saldin, E.A. Schneidmiller and M.V. Yurkov,
Nucl. Instrum. and Methods A490(2002)1.

\bibitem{ds2}
Heifets, G. Stupakov and S. Krinsky, Phys. Rev. ST Accel.Beams
5(2002)064401.

\bibitem{ds3}
Z. Huang and K.-J. Kim, Phys. Rev. ST Accel.Beams
5(2002)074401.

\bibitem{ultrf}
Jean-Claude Diels and W. Rudolph, Ultrafast Laser Phenomena,
Academic Press, San Diego, 1996.



\bibitem{gren}
P. O'Shea et al., Opt. Lett., vol 26, p. 932(2001).


\bibitem{czon}
P.L. Czonka, Part. Accel.8,225(1978).

\bibitem{book}
E.L. Saldin, E.A. Schneidmiller and M.V. Yurkov,
The physics of Free Electron Lasers, Springer,
Berlin-Heidelberg-New-York, 1999.


\bibitem{sem1}
B. E. Carlsten,
Nucl. Instrum. and Methods A285(1989)313.

\bibitem{sem2}
W. S. Graves et al.,
SLAC-PUB-9440 August 2002.

\bibitem{sem3}
D. H. Dowell et al.,
Nucl. Instrum. and Methods A507(2003)327.

\bibitem{sem4}
M. Krasilnikov et al., "Optimizing the PITZ Electron Source for VUV
FEL", presented at EPAC 2004, Lucerne, Switzerland, July 2004.

\end{thebibliography}
\end{document}